\documentclass[aps,pre,showpacs,floatfix,twocolumn,amsmath,amssymb]{revtex4-1}
\usepackage{color,graphicx,array,tabularx,bm,url,subcaption,dblfloatfix}
\newcolumntype{x}[1]{>{\centering\arraybackslash\hspace{0pt}}p{#1}}

\begin{document}

\title{Orientational order and topological defects in a dilute solutions of rodlike polymers at low Reynolds number}

\author{L. Puggioni}
\thanks{Corresponding author}
\email{leonardo.puggioni@unito.it}
\affiliation{Dipartimento di Fisica and INFN, Universit\'a degli Studi di Torino, via P. Giuria 1, 10125 Torino, Italy.}

\author{S. Musacchio}
\affiliation{Dipartimento di Fisica and INFN, Universit\`a degli Studi di
Torino, via P. Giuria 1, 10125 Torino, Italy.}

\begin{abstract}  
  The relationship between the polymer orientation and the chaotic flow,
  in a dilute solution of rigid rodlike polymers at low Reynolds number,
  is investigated by means of direct numerical simulations.
  It is found that the rods tend to align with the velocity field in order to minimize the friction with the solvent fluid, while regions of rotational disorder are related to strong vorticity gradients, and therefore to the chaotic flow. The "turbulent-like" behavior of the system is therefore associated with the emergence and interaction of topological defects of the mean director field, similarly to active nematic turbulence. The analysis has been carried out in both two and three spatial dimensions. 
\end{abstract}

\date{\today}

\maketitle

\section{Introduction}
\label{sec:intro}

The emergence of chaotic flows with a complex spatiotemporal structure at low Reynolds number ($Re$), \textit{i.e.} when fluid inertia is negligible with respect to viscous forces, is an intriguing phenomenon which occurs in various instances of complex fluids. A celebrated example is the regime of \textit{elastic turbulence},\cite{steinberg2021elastic,datta2022perspectives} which is observed in dilute solutions of flexible polymers (such as hydrolyzed polyacrylamide) at low $Re$ and large polymer elasticity.
\cite{groisman2000elastic,groisman2001efficient,groisman2004elastic,abed2016experimental,qin2017characterizing} 
Even though it occurs at low $Re$, this regime displays turbulent-like properties, such as increased resistance and efficient mixing, which have been successfully described by means of viscoelastic rheological models.\cite{fouxon2003spectra,berti2008two,grilli2013transition,ray2016elastic,plan2017lyapunov,van2018elastic,song2023self,singh2023intermittency}
A few years later a different instance of spontaneous chaotic flow was discovered in dense bacterial suspensions\cite{dombrowski2004self}: it was the first example of \textit{active turbulence}\cite{alert2022active}, subsequently observed, with different characteristics, in other typologies of active fluids.\cite{nishiguchi2015mesoscopic,sanchez2012spontaneous,creppy2015turbulence,blanch2018turbulent}

A distinctive feature of these turbulent-like flows at low $Re$ is the emergence of irregular velocity fluctuations in a broad range of scales, which typically manifest in a power-law spectral distribution of kinetic energy.
At variance with the process of turbulent energy cascade at high $Re$, in flows at low $Re$ the mechanisms which cause the spectral energy distribution depends on the details of the physical system.
In the case of elastic turbulence, the energy injected by external forcing at large scale
is redistributed at small scales by the elastic stress generated by polymers.\cite{fouxon2003spectra,steinberg2021elastic,song2023self}
In the case of active turbulence the energy is directly injected in a broad range of scales by the additional stress generated by the active particles (e.g., by the microswimmers).\cite{alert2022active,bowick2022symmetry}
In complex fluids with nematic symmetry the turbulent-like motion has been ascribed to the emergence and to the mutual interaction of topological defects of the average microscopic orientation (such as disclinations), generating a backflow in the coarse-grained velocity field.\cite{giomi2013defect,thampi2014instabilities,thampi2014vorticity,giomi2014defect,giomi2015geometry,vcopar2019topology}

Recently, a novel form of low-Reynolds chaotic flow has been observed in numerical simulations of a simple rheological model\cite{doi1988theory} of dilute solutions of rigid rod-like polymers (such as xanthan gum). \cite{emmanuel2017emergence,musacchio2018enhancement,puggioni2022enhancement}
Despite the fact that the stress generated by rigid polymers is substantially different from that of flexible elastic polymers, the properties of the resulting chaotic flow are qualitatively similar.
It particular previous studies~\cite{emmanuel2017emergence,musacchio2018enhancement,puggioni2022enhancement}
reported an increase in flow resistance and in mixing efficiency, which are observed in both two-dimensional (2D) and three-dimensional (3D) simulations of the model. 

In the present study, we pursue the investigation of the chaotic regime in dilute solutions of rigid rod-like polymers,
focusing on the interplay between the solvent velocity field and the orientational order of the polymer phase.
We present a detailed study of the statistics of the polymer orientation.
The latter is described in terms of the the Westin coefficients\cite{callan2006simulation},
which are commonly used in nematic liquid crystals theory.
Our results show that the chaotic flow is produced by the emergence of topological defects
in the average polymer orientation.
This process is qualitatively reminiscent the phenomenon observed in nematic active fluids\cite{giomi2013defect,thampi2014instabilities,thampi2014vorticity,giomi2014defect,giomi2015geometry,vcopar2019topology}. 

The rest of the paper is organized as follows. In Sec.~\ref{sec:2} we present the Doi-Edwards model for the dynamics of dilute rodlike polymers solutions, we provide a brief description of the Westin coefficients 
and we discuss the details of the numerical simulations.
In Sec.~\ref{sec:res} we present the results concerning the statistics of the orientation of the rods and 
its relationship with the chaotic flow. Finally, we present a comparison between the results
of numerical simulation of the model in 2D and 3D simulations.
Sec.~\ref{sec:con} is devoted to conclusions.

\section{The Doi-Edwards model}
\label{sec:2}
The Eulerian model for a dilute solution of inertialess rodlike polymers, at rotational equilibrium with the solvent fluid, was formulated by Doi and Edwards\cite{doi1988theory}.
The direction and the strength of the local alignment of the polymer phase is described by the configuration tensor field $R_{ij} \left( \bm{x},t \right) = \langle a_i a_j \rangle_{\mathcal{V}}$, where $\bm{a}$ is the orientation vector of an individual rod and the average is taken over an infinitesimal volume element $\mathcal{V}$, at position $\bm{x}$
and time $t$. The configuration tensor is symmetric and has unit trace. 
Assuming that the polymers are immersed in an incompressible velocity field $\bm{u} \left( \bm{x}, t \right)$, the dynamics of the system is given by the following equations:
\begin{subequations}
\begin{align}
\partial_t {u_i} + {u_k}\partial_k{u_i} =& -\partial_i p + \nu \partial^2{u_i} + 
\partial_k{\sigma_{ik}} + {f_i}\,,\label{eq:sys1a}
\\
\partial_t R_{ij} + {u_k} \partial_k R_{ij} =& (\partial_k{u_i}) R_{kj}
 + R_{ik}(\partial_k{u_j}) \nonumber \\ & -2 R_{ij}(\partial_l u_k) R_{kl},
\label{eq:sys1b}
\end{align}
\label{eq1}
\end{subequations}
where $\nu$ is the kinematic viscosity of the solvent fluid and $\bm{f}$ is the external body-force which sustains the flow. The pressure field $p\left(\bm{x}, t \right)$ preserves the incompressibility constraint $\partial_k u_k =0$. The polymer stress $\sigma_{ij}$ takes the form\cite{doi1988theory}: 
\begin{equation}\label{eq2}
\sigma_{ij} = 6 \nu \eta_p R_{ij} \left( \partial_l u_k \right) R_{kl}.
\end{equation}
The intensity of the polymer feedback to the flow is determined by the parameter $\eta_p$, taking into account the geometrical properties of the rods and their volume fraction. In the case of an aqueous solution of xantham gum, an experimental relationship $\eta_p = 0.011147 C^{1.422}$ was found, where $C$ is the concentration in weight parts per million (wppm)\cite{amarouchene2008reynolds}. The expression \eqref{eq2} is based on a simple quadratic
approximation, although more sophisticated closures have been proposed.\cite{advani1987use,chung2002invariant,gillissen2007performance,montgomery2011exact} It has been numerically shown that the model \eqref{eq1} is able to reproduce, for large values of the Reynolds number, the phenomenon of turbulent drag reduction in wall-bounded flows \cite{benzi2005additive,ching2006turbulent,procaccia2008colloquium,amarouchene2008reynolds,benzi2008comparison}.
The model \eqref{eq1} also contains terms related to the Brownian rotations of the rods, but these effects can be safely disregarded in chaotic flows in which the rotational dynamics is dominated by the velocity gradients.\cite{gillissen2007stress,puggioni2022enhancement}

It is worth to remark that Eq. \eqref{eq:sys1b} is analogous to the equation for the evolution of the nematic order parameter in liquid crystals, with a free energy equal to zero.\cite{de1993physics,beris1994thermodynamics}
In particular, the configuration tensor $R_{ij}$ is related to the nematic traceless tensor $Q_{ij}$
by $Q_{ij} = R_{ij} - \left(1/D \right) \delta_{ij}$,
where $D$ is the number of spatial dimensions and $\delta_{ij}$ the standard Kronecker delta. 
Nonetheless, the stress associated with liquid crystals is very different from the polymer stress \eqref{eq2}. 

\subsection{Degree of order and Westin coefficients.}
The nematic symmetry of the polymer phase allows us to adopt concepts from liquid crystals theory
for the quantification of its degree of order. 
Since $R_{ij}$ is a real-valued, symmetric unit-trace tensor,
it possesses non-negative real eigenvalues $\lambda_1$, $\lambda_2$
and $\lambda_3$ with $1 \geq \lambda_1 \geq \lambda_2 \geq \lambda_3 \geq 0$,
and $\text{Tr} \bm{R}=\lambda_1+\lambda_2+\lambda_3 = 1$.
They are associated with the three eigenvectors $\bm{n}$, $\bm{m}$ and $\bm{l}$, respectively.
From the eigenvalues we can define the three Westin coefficients\cite{callan2006simulation}:
\begin{equation}
c_l = \lambda_1-\lambda_2, \qquad c_p = 2\left( \lambda_2- \lambda_3 \right), \qquad c_s = 3 \lambda_3.
\end{equation}
The value $c_l\approx 1$ corresponds to local uniaxial order,
\textit{i.e.} the rods are oriented in the direction given by the first eigenvector $\bm{n}$.
The value $c_p \approx 1$ corresponds to local biaxial order,
\textit{i.e.} the rods are randomly oriented in the plane defined by the first two eigenvectors $\bm{n}$ and $\bm{m}$.
Finally, $c_s \approx 1$ corresponds to local isotropy,
\textit{i.e.} the rods are randomly oriented in the three-dimensional space.

The identification of topological defects in three-dimensional liquid crystals
by means of Westin coefficients is well established.\cite{lopez2012defect,wand2015monte,chiccoli2018defect,negro2023topological}
In the case of rod-like polymers, the topological defects in the orientation of the director field $\bm{n}$
can be defined in terms of the Westin coefficients as the regions in which $c_l < c_p$ (defects with biaxial structure)
or $c_l < c_s$ (defects with isotropic structure).
Recalling that the nematic traceless tensor $Q_{ij}$ can be expressed
in terms of the scalar order parameters $S$ and $\epsilon$ as:
\begin{equation}
\bm{Q} = S  \left( \bm{n} \bm{n} - \frac{1}{3} \bm{I}\right) + \epsilon \left( \bm{m} \bm{m} - \frac{1}{3} \bm{I}\right),
\end{equation}
and that $\bm{Q} = \bm {R} - \left(1/3 \right) \bm{I}$, 
the Westin coefficients are related to $S$ and $\epsilon$ as follows:\cite{callan2006simulation}
\begin{equation}
c_l = S-\epsilon, \qquad c_p = 4\epsilon, \qquad c_s = 1 - S- 3 \epsilon.
\end{equation}
In the two-dimensional case, we simply have $c_l=S=\lambda_1-\lambda_2$.

\subsection{Numerical methods and configuration}
In the numerical simulations of the model \eqref{eq1}
the flow is sustained by an external monochromatic force 
$\bm{f} \left( \bm{x} \right) = \left[ F \cos \left( Kz \right),0,0 \right]$,
where $F$ is the amplitude and $K$ is the wavenumber of the forcing.
In the absence of polymers, Eq.~\eqref{eq:sys1a} with this forcing 
admits the laminar solution $\bm{u} \left( \bm{x} \right)= \left[ U_0 \cos \left( Kz \right),0,0 \right]$
with amplitude $U_0 = F/(K^2 \nu)$ (the so-called \textit{Kolmogorov flow}).
The laminar solution is linearly stable when the Reynolds number $Re=U_0/(K\nu)$
is smaller than the critical value $Re_c = \sqrt{2}$.\cite{meshalkin1961investigation}

A remarkable feature of the Kolmogorov flow is that the mean velocity profile $\overline{\bm{u}}(z)$,
defined as the average of the velocity field over the $x$ and $y$ coordinates and over time $t$,
maintains the symmetry of the external forcing also in turbulent or chaotic regimes, i.e.,
$\overline{\bm{u}}(z) = \left[ U \cos \left( Kz \right),0,0 \right]$. 
The amplitude $U$ of the mean flow allows to define 
a drag coefficient $f = F/(KU^2)$\cite{musacchio2014turbulent}
which has been used to investigate the asymptotic behavior of the turbulent drag
in Newtonian flow at large $Re$\cite{borue1996numerical,musacchio2014turbulent}
and the phenomenon of drag reduction in non-Newtonian flows\cite{boffetta2005drag,sozza2020drag}. 
In the low $Re$ regime, this flow configuration has been adopted to study the phenomena of 
elastic turbulence\cite{berti2008two}, elastic waves\cite{berti2010elastic}
and other forms of instabilities\cite{boi2013minimal}.
Here, we exploit the symmetries of the mean flow to investigate the differences between the 
statistics of the polymer orientation in shear-dominated and flow-dominated regions. 

The numerical integration of Eqs.~(\ref{eq:sys1a},\ref{eq:sys1b}) with Kolmogorov forcing
has been performed by means of a standard dealiased pseudospectral method
in a triple-periodic cubic domain of size $L=2\pi$, uniformly discretized in $N^3=256^3$ gridpoints.
Time integration is carried out using a fourth-order Runge-Kutta scheme
with implicit integration of linear terms.
In order to assure numerical stability, Eq.~\eqref{eq:sys1b} is supplemented with
a diffusive term $\kappa \partial^2 R_{ij}$.\cite{sureshkumar1995effect}

We performed four sets of simulations varying the parameter $\eta_p \in \{ 5,6,7,8\}$.
The values of the other parameters of the model are kept fixed as $\nu=1$, $\kappa = 4\times 10^{-3}$, $K=4$, $F=64$. 
We also performed a further ensemble of simulations
in a two-dimensional squared domain with $512^2$ gridpoints and with identical parameters,
which allows to investigate the dependence of the statistics on the dimensionality of the system.
For this purpose, the values of the parameter $\eta_p$ in the 2D simulations have been rescaled
according to the dimensional relation $\eta_p^{2D} = (2/3) \eta_p^{3D}$ proposed in \cite{puggioni2022enhancement}.

In all the simulations, the velocity field is initialized with the the laminar solution
$\bm{u} \left(\bm{x},t=0 \right)= \left[ U_0 \cos \left( Kz \right),0,0 \right]$. 
The polymer configuration tensor is initialized with an uniaxial configuration $R_{ij}(\bm{x},t=0) = n_in_j$ 
with random orientation of the director $\bm{n}$ at each point $\bm{x}$.
For each value of of $\eta_p$ we realized three independent simulations
with different initial random orientation. 
After an initial transient, the system reaches a
statistically stationary chaotic state.\cite{emmanuel2017emergence,puggioni2022enhancement}
The results presented henceforth have been obtained in the stationary chaotic regime. 


\section{Results}
\label{sec:res}

\begin{figure}[h]
	\centering
	\includegraphics[width=0.82\linewidth]{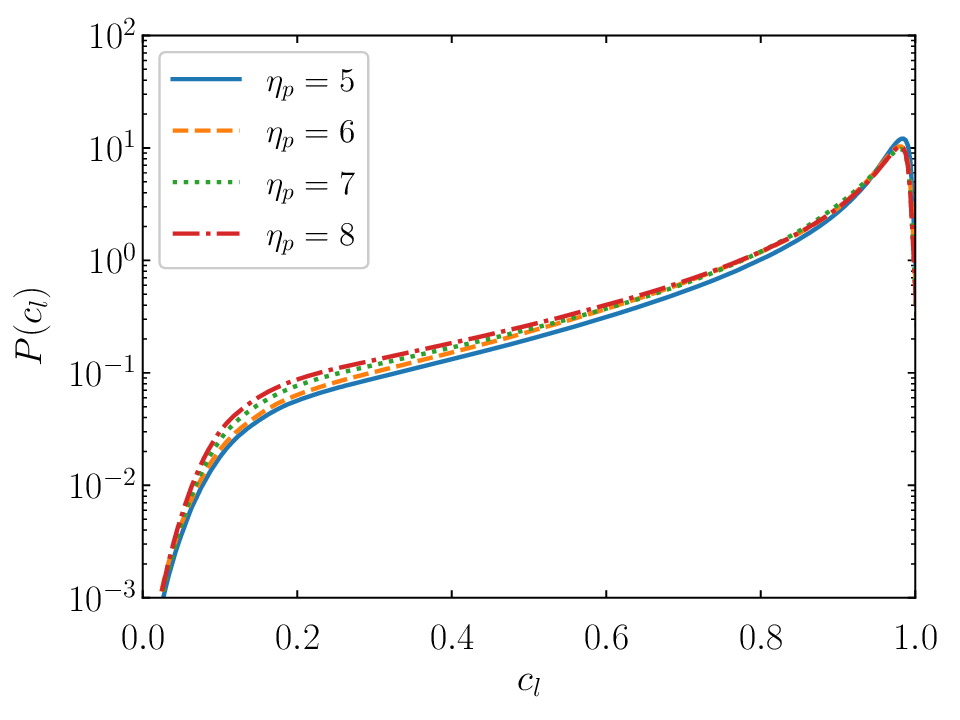}
	\includegraphics[width=0.82\linewidth]{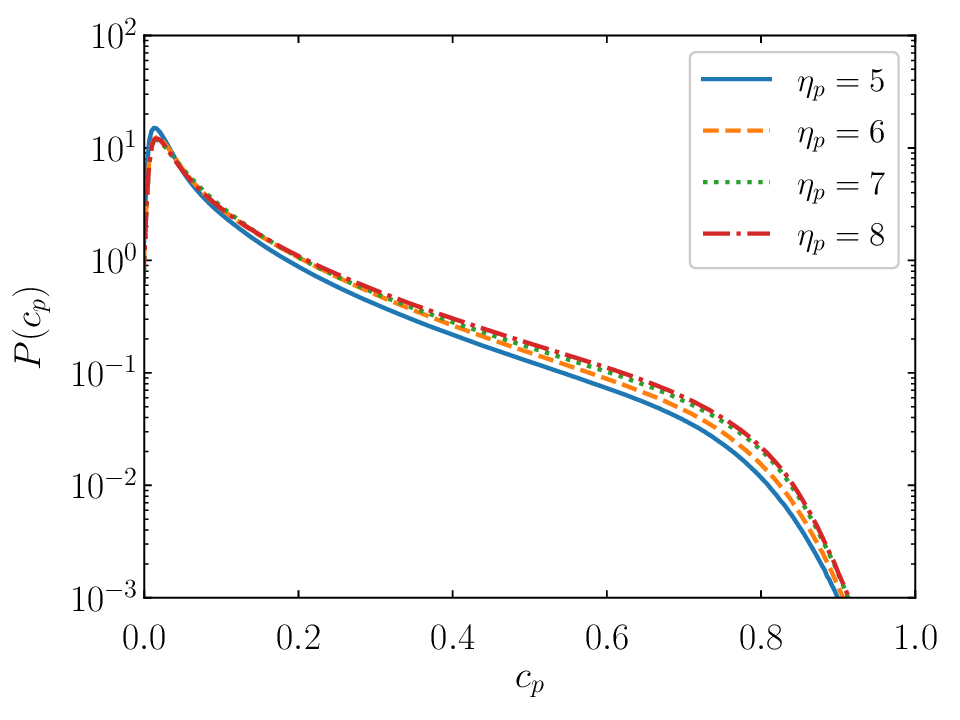}
	\includegraphics[width=0.82\linewidth]{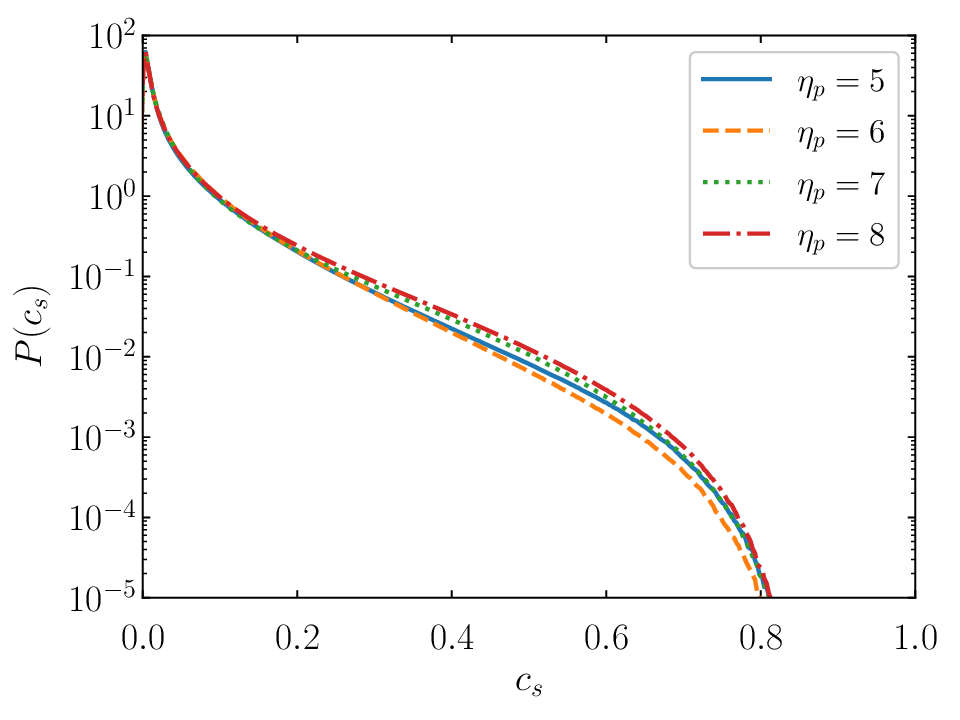}
	\caption{Probability density functions of the Westin coefficients
          $c_l$ (top), $c_p$ (center) and $c_s$ (bottom)
          for different values of concentration coefficient $\eta_p$.
	}
	\label{fig1}
\end{figure}

\subsection{Statistics of polymer orientation and correlations with the mean flow}
A first information about the statistics of polymer orientation is
given by the probability density functions (pdf) of the three Westin coefficients (Fig. \ref{fig1}).
We observe that $P(c_l)$ is peaked around $c_l\approx 1$ (Fig. \ref{fig1} top),
while $P(c_p)$ and $P(c_s)$ are peaked around $0$. This shows that the large majority of points displays an uniaxial configuration.
We also note that the right tail of $P(c_s)$  (Fig. \ref{fig1} bottom) decays much faster than that of $P(c_p)$ (Fig. \ref{fig1} center), 
meaning that there are significantly more points with biaxial order than the points with isotropic configuration.

\begin{figure}[h]
	\centering
	\includegraphics[width=0.98\linewidth]{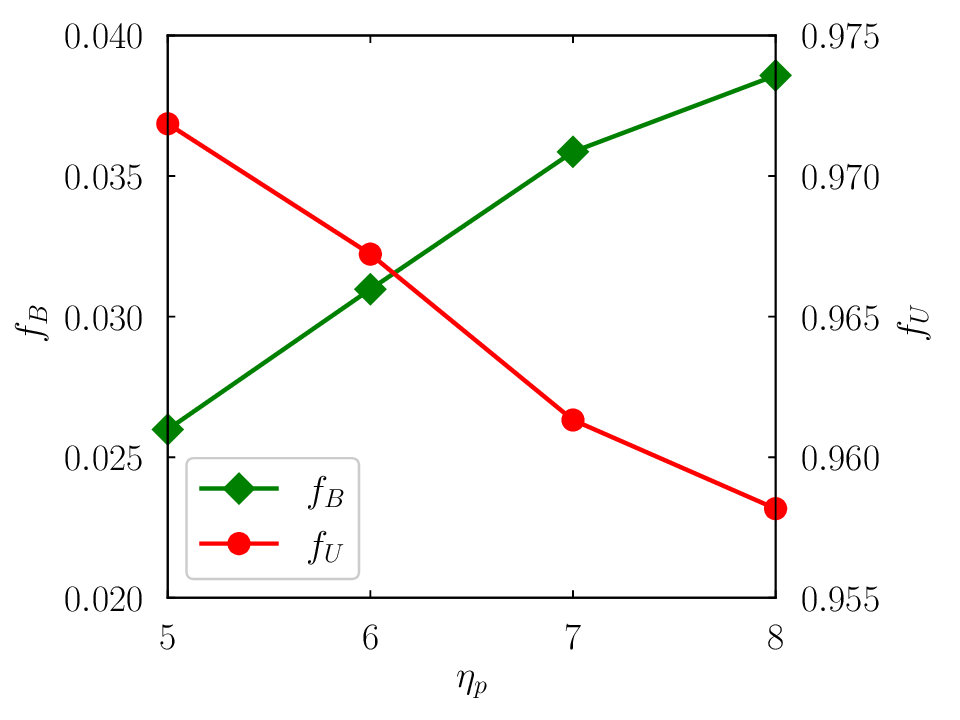}
	\caption{Fraction of points with prevalent
          uniaxial order $f_U$ (red circles) 
          and with prevalent biaxial order $f_S$ (green diamonds)
          as a function of the concentration $\eta_p$.
	}
	\label{fig2}
\end{figure}

The dependence on the concentration parameter $\eta_p$ is evident in Fig.~\ref{fig2},
which shows the fraction $f_U$ of points with prevalent uniaxial order (i.e., points where $c_l > c_p, c_s$)
and the fraction $f_B$ of points with prevalent biaxial order (i.e., points where $c_p > c_l, c_s$). 
We observe a relevant increase of $f_B$ from $2.5\%$ to $4\%$ at increasing $\eta_p$,
accompanied by a decrease of $f_U \simeq 1 - f_B$.
The fraction of points with prevalent isotropic configuration is much smaller (not shown). 
This result constitutes a first indication that the topological defects of the director field
$\bm{n}(\bm{x},t)$, which are defined as the region in which $c_l < max(c_p,c_s)$, 
are related to the "turbulent-like" behavior observed in the velocity field.
The observation that the biaxial configuration for the defects is much more likely then the isotropic one 
is reminiscent of what happens in nematic liquid crystals,
in which the core of a wedge-like disclination possesses a biaxial structure.\cite{schopohl1987defect}

\begin{figure}[t]
	\centering
	\includegraphics[width=0.82\linewidth]{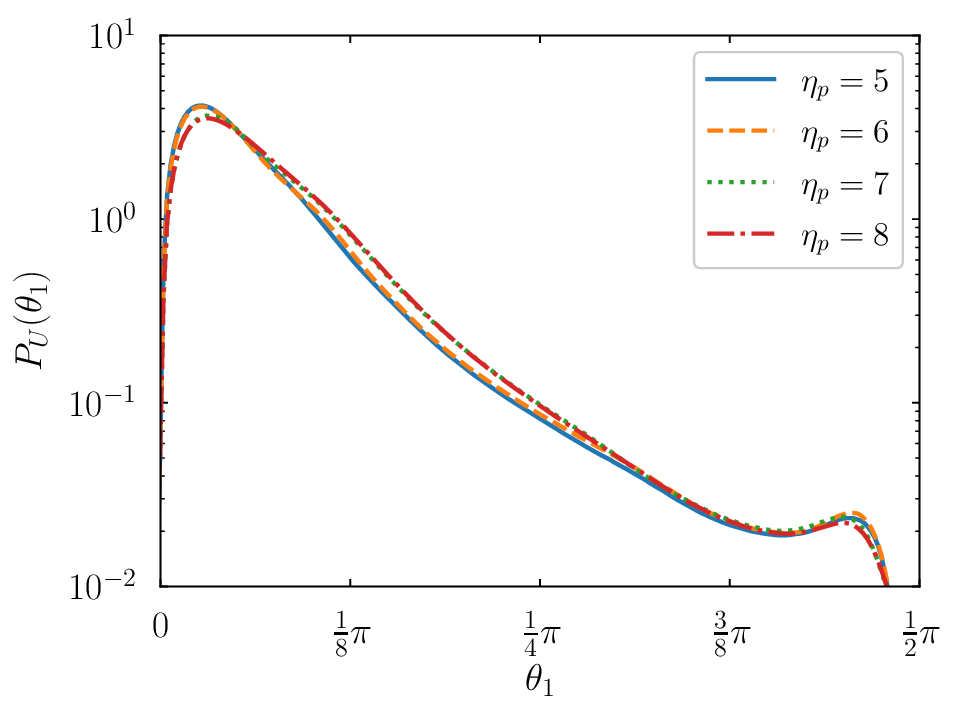}
	\includegraphics[width=0.82\linewidth]{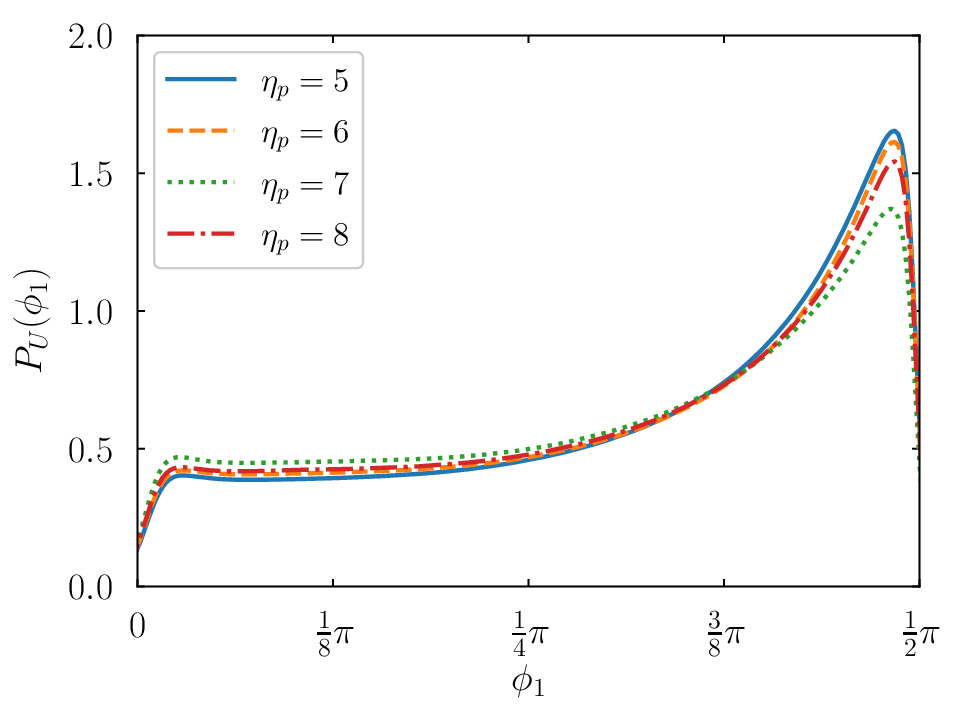}
	\includegraphics[width=0.82\linewidth]{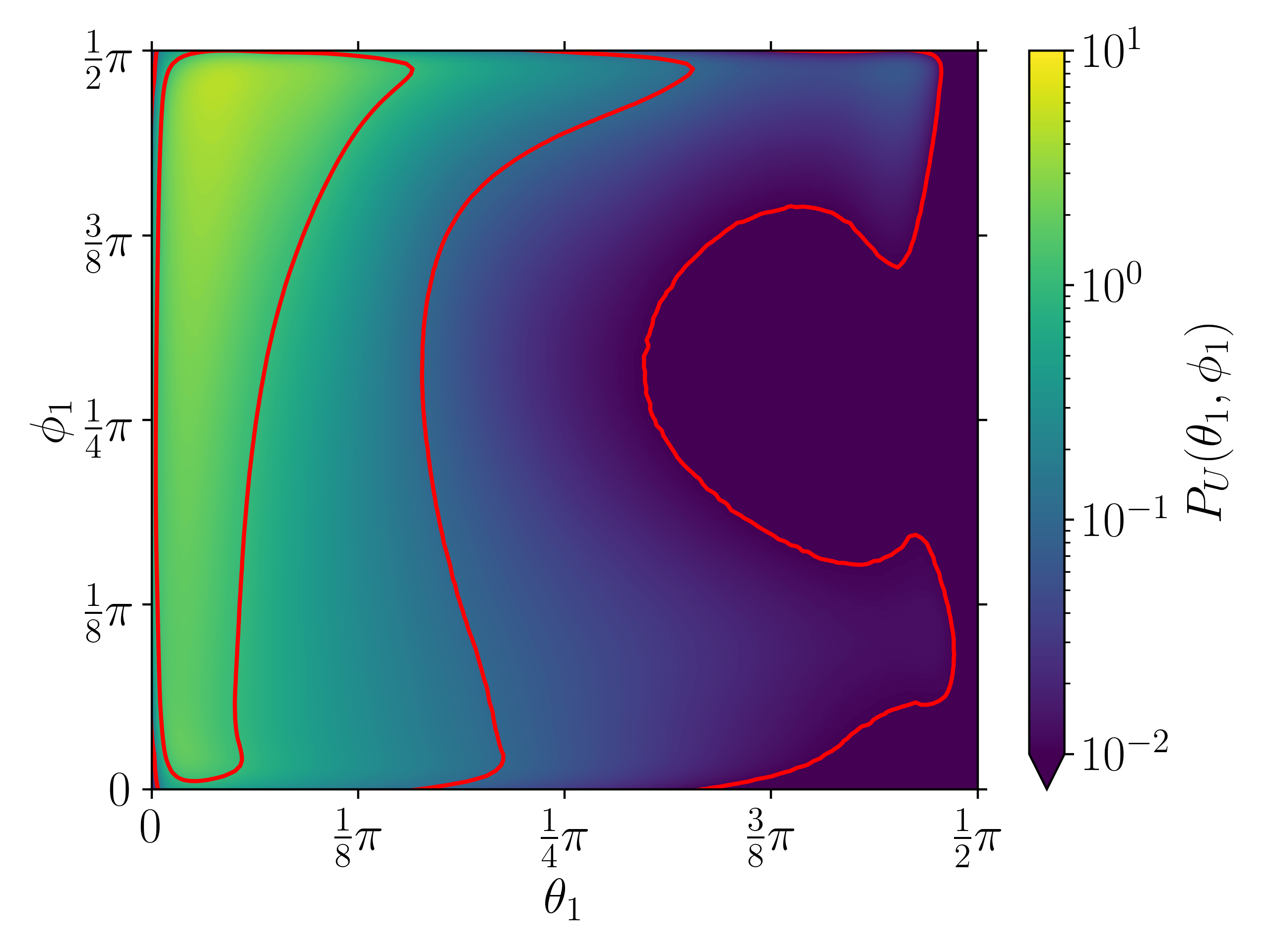}
	\caption{Pdfs of the angle $\theta_1$ between the director $\bm{n}$ and the $x$-axis (top),
          and of the angle $\phi_1$ representing the orientation of the projection of $\bm{n}$ in the $y$-$z$ plane (center), 
          for different values of concentration coefficient $\eta_p$.
          Bottom: joint pdf of the orientation of $\bm{n}$ in the $\theta_1$-$\phi_1$ plane
          averaging on all the values of concentration coefficient $\eta_p$.
          The statistics is conditioned to regions with prevalent uniaxial configuration ($c_l > c_p,c_s$)
	}
	\label{fig3}
\end{figure}
\begin{figure}[t]
	\centering
	\includegraphics[width=0.82\linewidth]{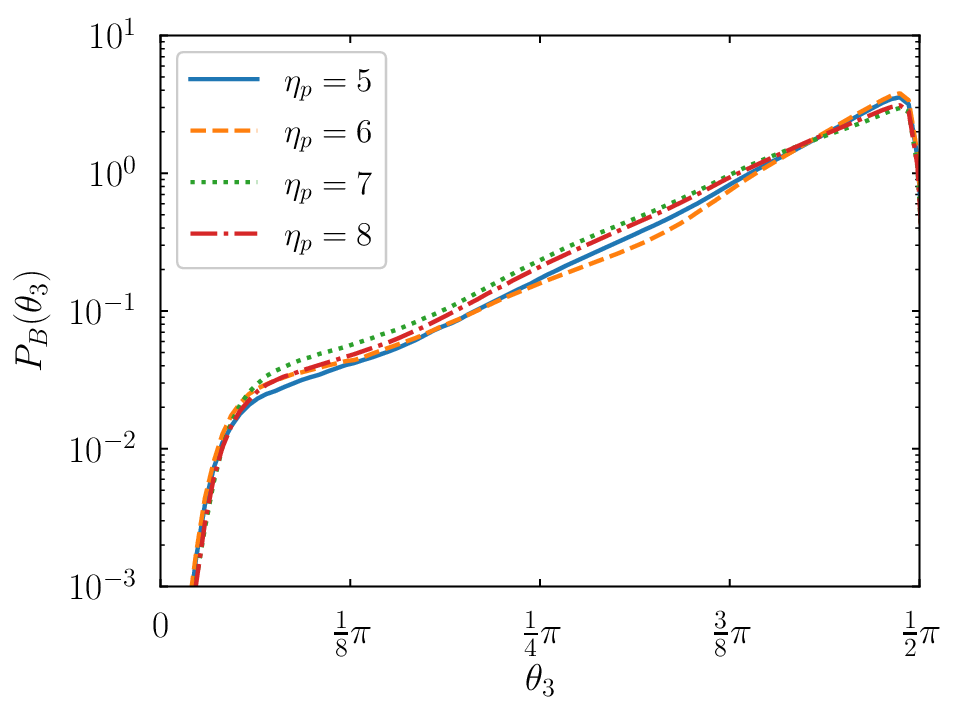}
	\includegraphics[width=0.82\linewidth]{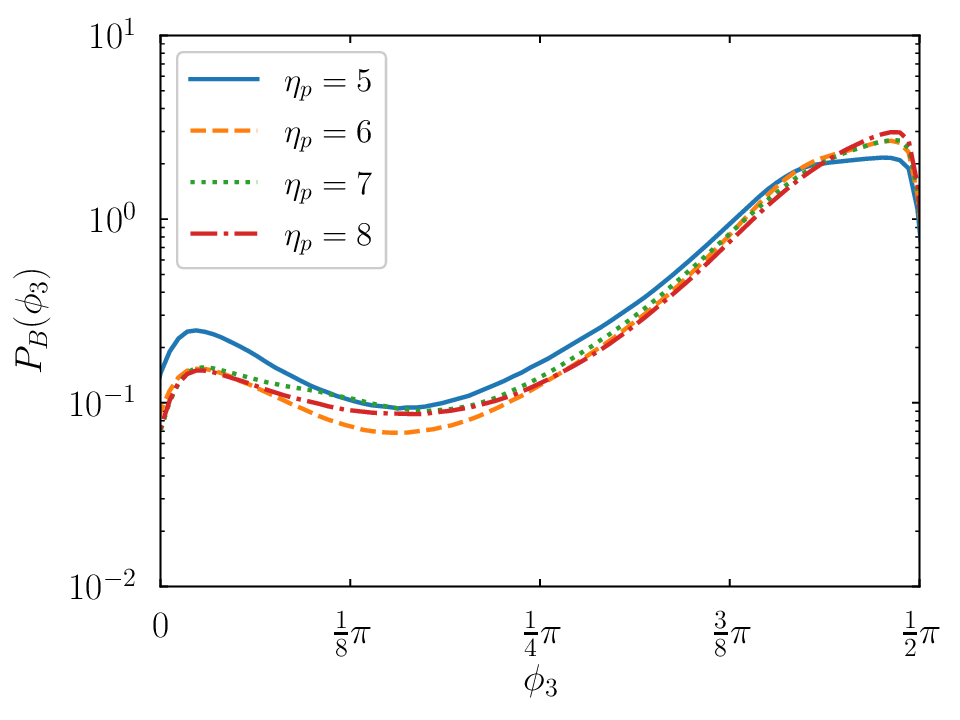}
	\includegraphics[width=0.82\linewidth]{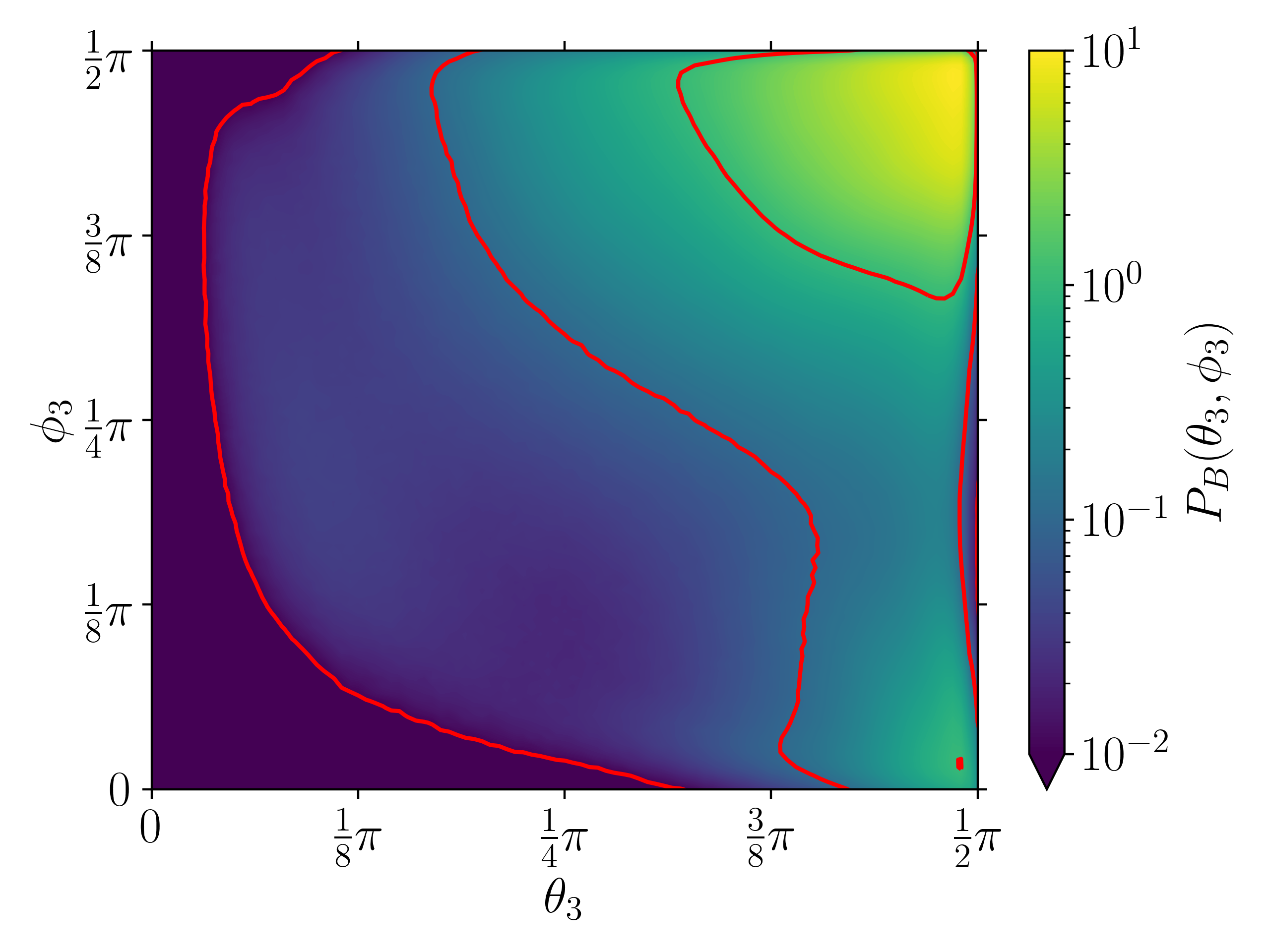}
	\caption{Top and center: pdfs of the angle $\theta_3$ between the third eigenvector $\bm{l}$ and the $x$-axis (top),
          and of the angle $\phi_3$ representing the orientation of the projection of $\bm{l}$ in the $y$-$z$ plane (center), for different values of concentration coefficient $\eta_p$. Bottom: joint pdf of the orientation of $\bm{l}$ in the $\theta_3$-$\phi_3$ plane, averaging on all the values of concentration coefficient $\eta_p$. The statistics is conditioned to regions with prevalent biaxial configuration ($c_p > c_l,c_s$)
	}
	\label{fig4}
\end{figure}

In the regions in which the director $\bm{n}$ is unambiguously defined (\textit{i.e.} where $c_l > c_p,c_s$), 
the statistics of the orientation of $\bm{n}$
can be investigated in terms of the angles $\theta_1 = \arccos |n_x|$ and $\phi_1=\arccos\left( |n_y|/\sqrt{n^2_y+n^2_z} \right)$.
The angle $\theta_1$ represents the angle between $\bm{n}$ and the mean flow
which is oriented along the $x$-axis (the streamwise direction). 
The angle $\phi_1$ represents the orientation of the projection of $\bm{n}$ in the $y$-$z$ plane.
The value $\phi_1=0$ corresponds to the alignment with the $y$-axis (the neutral, or spanwise, direction)
while $\phi_1=\pi/2$ indicates the alignment with the $z$-axis (the mean shear, or cross-flow, direction).
The pdfs of $\theta_1$ and $\phi_1$
conditioned to the points with prevalent uniaxial order ($c_l > c_p,c_s$) are shown in Fig. \ref{fig3}.
We find that $P_U(\theta_1)$ is peaked at small angles $\theta \approx \pi/40$, indicating that 
$\bm{n}$ tends to be almost (although not completely) parallel to the mean flow. 
The deviation from the $x$-axis in the $y-z$ plane shows a preferential orientation
in the direction the mean shear (the $z$-axis) since $P_U(\phi_1)$ is peaked at $\phi\sim\pi/2$.
Nonetheless, the probability of other values of $\phi_1$ is not negligible
and it is almost constant for $\phi_1 < \pi/4$.
The joint pdf $P_U\left( \theta_1,\phi_1 \right)$ (averaged on all the values of $\eta_p$)
confirms that the rods tend to align with the mean flow in the regions where the director $\bm{n}$
is unambiguously defined (\textit{i.e.} where $c_l > c_p,c_s$). 
The weak preference for the orientation along the $z$ axis compared to $y$ axis
is presumably due to the fact that velocity fluctuations
in the cross-flow direction are more intense than in the spanwise direction.\cite{puggioni2022enhancement}

In the regions with prevalent biaxial order (\textit{i.e.} where $c_p > c_l,c_s$)
the first two eigenvectors $\bm{n}$ and $\bm{m}$ are not unambiguously defined.
In this case, it is convenient to characterize the statistics of polymer orientation
by means the third eigenvector $\bm{l}$, which is perpendicular the plan defined by $\bm{n}$ and $\bm{m}$.
The orientation of $\bm{l}$ is described by the angles $\theta_3 = \arccos | l_x|$
and $\phi_3=\arccos\left( |l_y|/\sqrt{l^2_y+l^2_z} \right)$. 
In Figure \ref{fig4} we show pdfs of $\theta_3$ and $\phi_3$,
as well as the joint pdf $P_B(\theta_3,\phi_3)$ averaged on all the values of $\eta_p$.
The statistics is conditioned to biaxial regions $c_p > c_l,c_s$. 
The pdf $P_B(\theta_3)$ is peaked at $\theta_3 \lesssim \pi/2$,
with an exponential tail $P_B(\theta_3) \sim A e^{\theta_3}$ in the range $\pi/15 \lesssim \theta \lesssim \pi/2$,
while $p(\phi_3)$ has a more complex shape, with a maximum at $\phi_3$ close to $\pi/2$ and a secondary peak close to $0$.
The two-dimensional joint pdf $P_B(\theta_3,\phi_3)$ averaged on all values of $\eta_p$
has a clear maximum close to the corner $\{\theta_3=\pi/2, \phi_3 = \pi/2 \}$ (Fig.~\ref{fig4}, bottom),
which corresponds to a prevalent direction of $\bm{l}$ aligned with the $z$-axis. 
This shows that inside the biaxial regions the rods are mostly oriented in the $x$-$y$ plane,
which is orthogonal to the direction of the mean shear.
We also find a secondary peak in the joint pdf $P_B(\theta_3,\phi_3)$
close to the corner $\{\theta_3=\pi/2, \phi_3 = 0 \}$,
corresponding to rods oriented in a plane almost parallel to the $x$-$z$ plane.

Recalling that the regions with prevalent biaxial order correspond
to topological defects of the orientation of the director field $\bm{n}$, 
the results obtained so far can be summarized as follows:
In the regions characterized by nematic order the polymers are mostly aligned in the direction of the mean flow;
conversely, within the topological defects they are preferentially oriented in the plane orthogonal to the mean shear. 

\begin{figure}[h]
	\centering
	\includegraphics[width=0.99\linewidth]{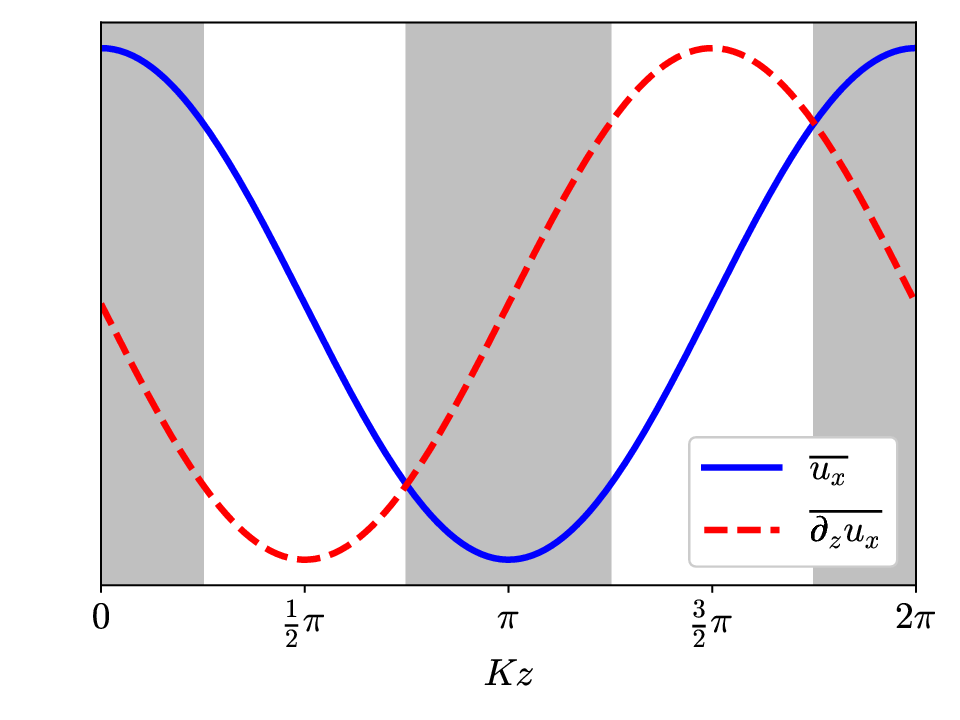}
	\caption{Symbolic sketch representing the partition of the spatial
          domain along the $z$-direction into flow-dominated regions (gray background)
          and shear-dominated regions (white background).
	}
	\label{fig5}
\end{figure}

The influences of the structure of the mean flow and of the mean shear on the statistics of polymer orientation can be further investigated
by dividing the spatial domain into $flow$-dominated and $shear$-dominated regions. 
The Kolmogorov flow is an optimal setup for this purpose, since it allows to identify very easily these regions
thanks to the property that the vertical profiles of the mean velocity $\overline{u_x}(z)$
and of the mean velocity gradient $\overline{\partial_z u_x}(z)$
remain monochromatic also in the chaotic regime\cite{emmanuel2017emergence,puggioni2022enhancement}:  
\begin{equation}
\overline{u_x}(z) = U\cos \left( Kz \right), \qquad \overline{\partial_z u_x}(z) = -KU \sin \left( Kz \right). 
\end{equation}
Here and in the following the overbar $\overline{ \left[ \cdot \right] } $ indicates the average over $x$ and $y$ coordinates and time $t$.  
The $flow$-dominated regions (denoted as $F$) are defined as the regions located around the extremals of the mean flow, i.e., 
$Kz \in \left[ 0, \pi/4 \right) \cup \left[ 3\pi/4, 5\pi/4 \right) \cup \left[ 7\pi/4 , 2\pi \right]$
    (gray regions in Figure \ref{fig5}).    
The $shear$-dominated regions (denoted as $S$) are defined as the regions located around the extremals of the mean shear, i.e., 
$Kz \in \left[ \pi/4, 3\pi/4\right) \cup \left[ 5\pi /4, 7\pi/4 \right)$
    (white regions in Figure \ref{fig5}).
    
\begin{figure}[t]
	\centering
	\includegraphics[width=0.83\linewidth]{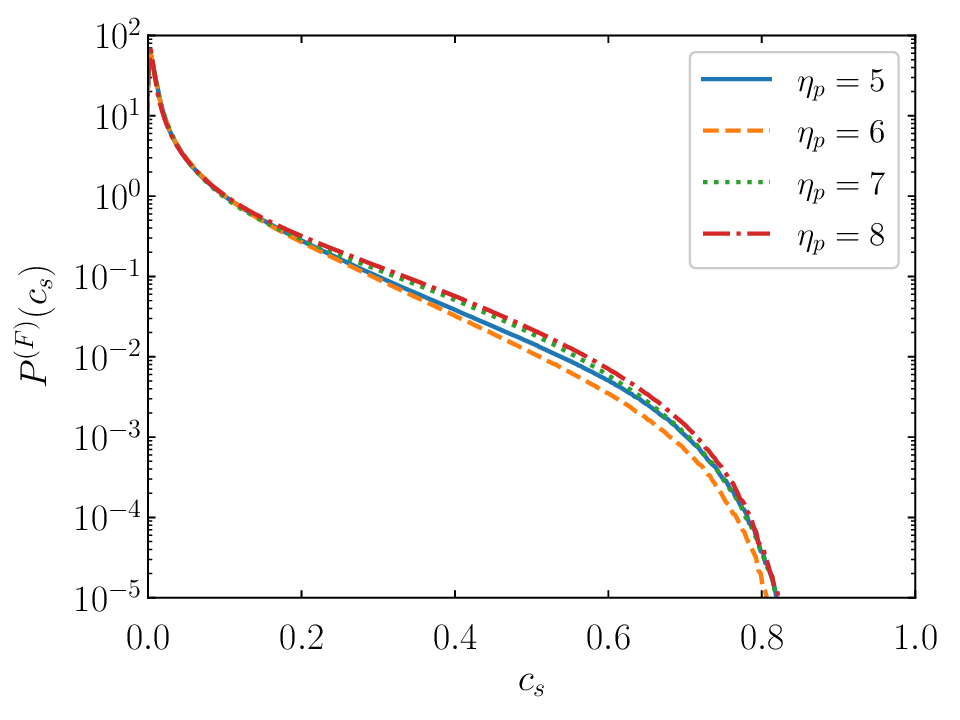}
	\quad
	\includegraphics[width=0.83\linewidth]{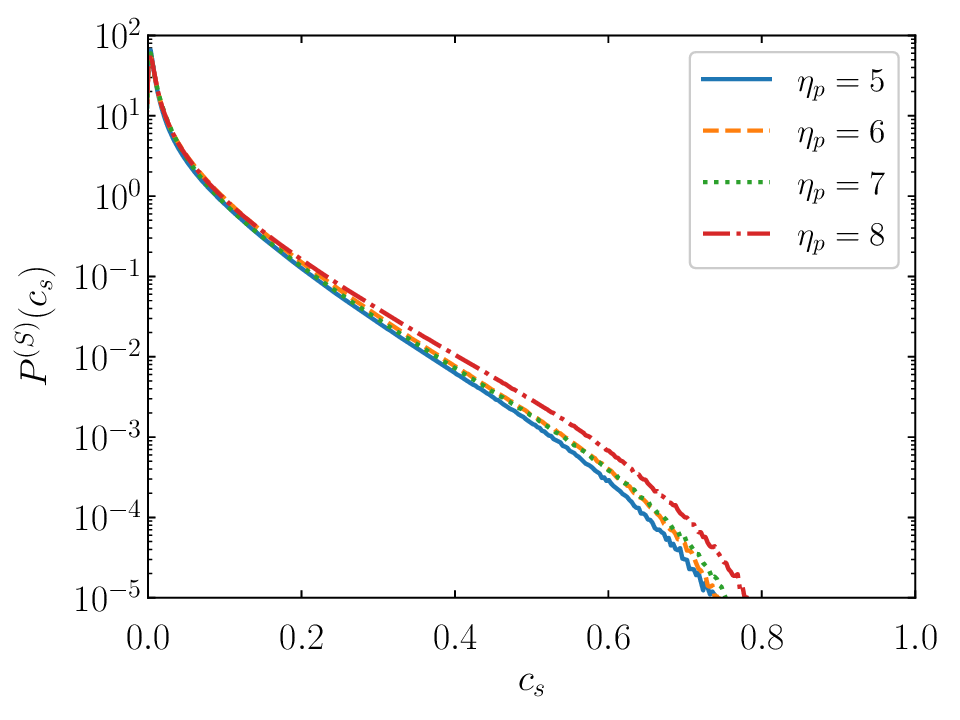}
	\caption{ Pdfs of isotropy coefficient $c_s$
          conditioned to flow- (top) and shear- (bottom) dominated regions,
          for different values of concentration coefficient $\eta_p$.
	}
	\label{fig6}
\end{figure}
\begin{figure}[t]
	\centering
	\includegraphics[width=0.85\linewidth]{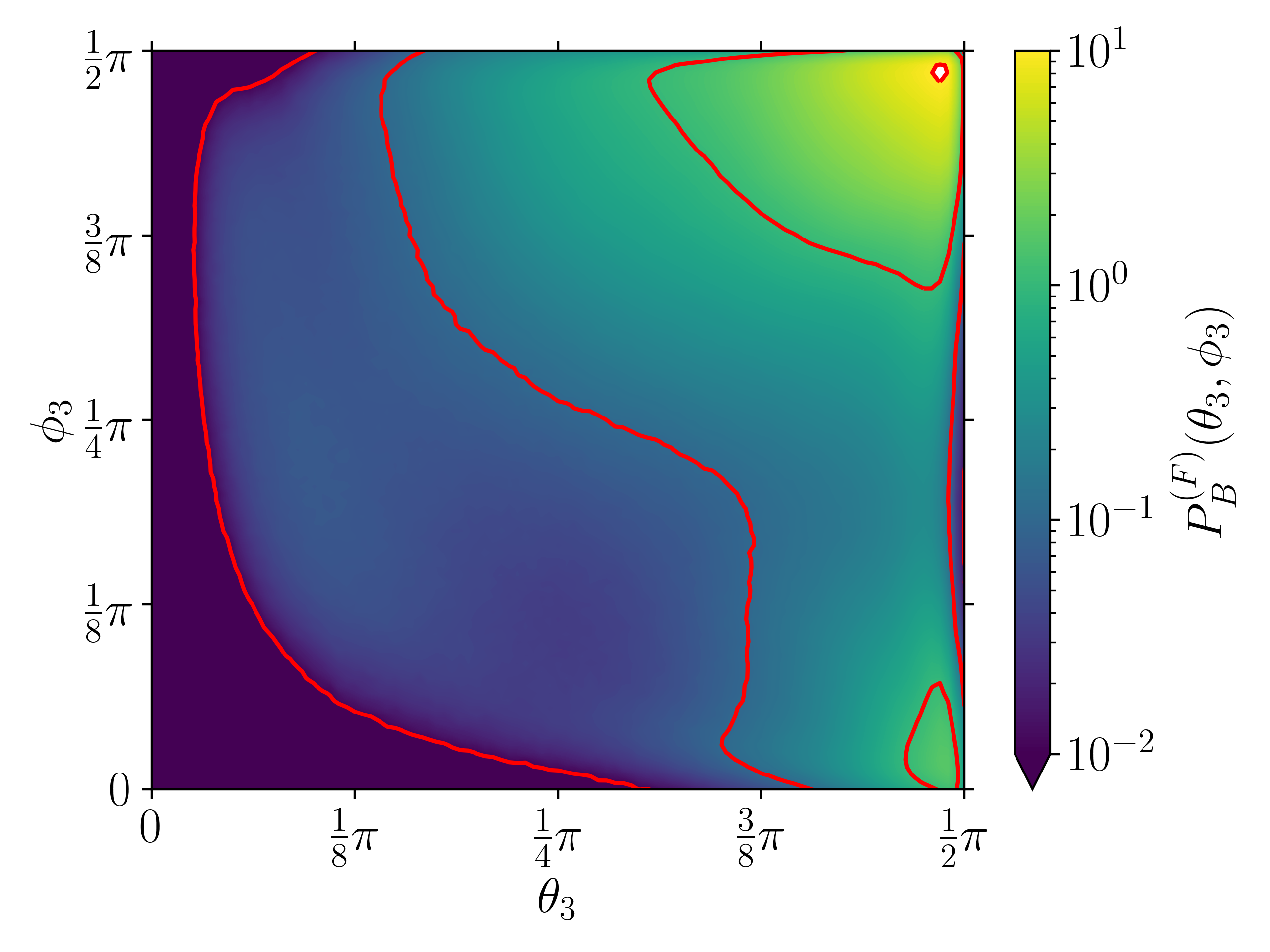}
	\includegraphics[width=0.85\linewidth]{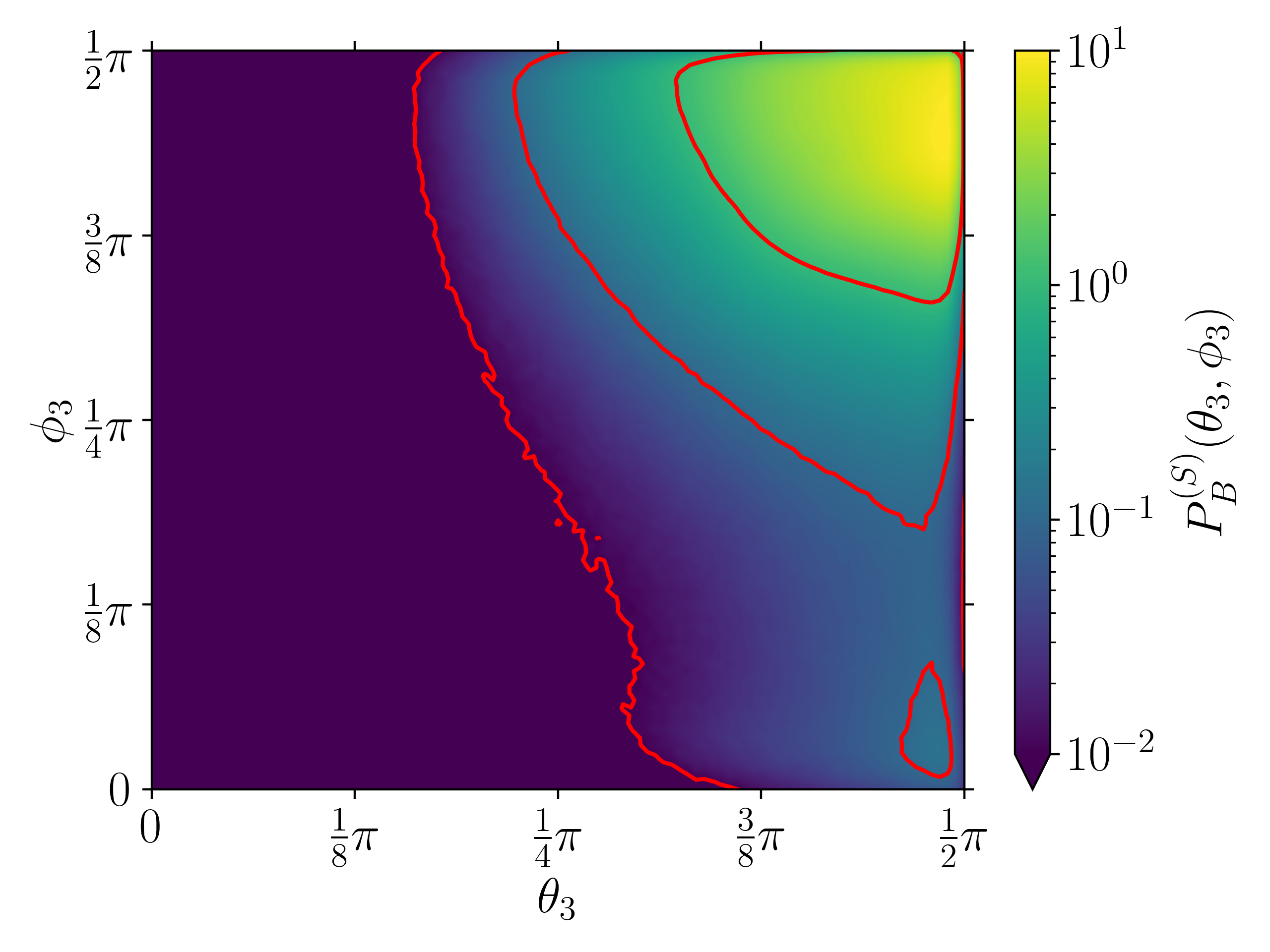}
	\caption{ Joint pdfs of the third eigenvector $\bm{l}$
          in the $\theta_3$-$\phi_3$ plane,
          conditioned to flow- (top) and shear- (bottom) dominated regions.
          The statistics is averaged on all the values of concentration $\eta_p$.
	}
	\label{fig7}
\end{figure}
\begin{figure}[t]
	\centering
	\includegraphics[width=0.82\linewidth]{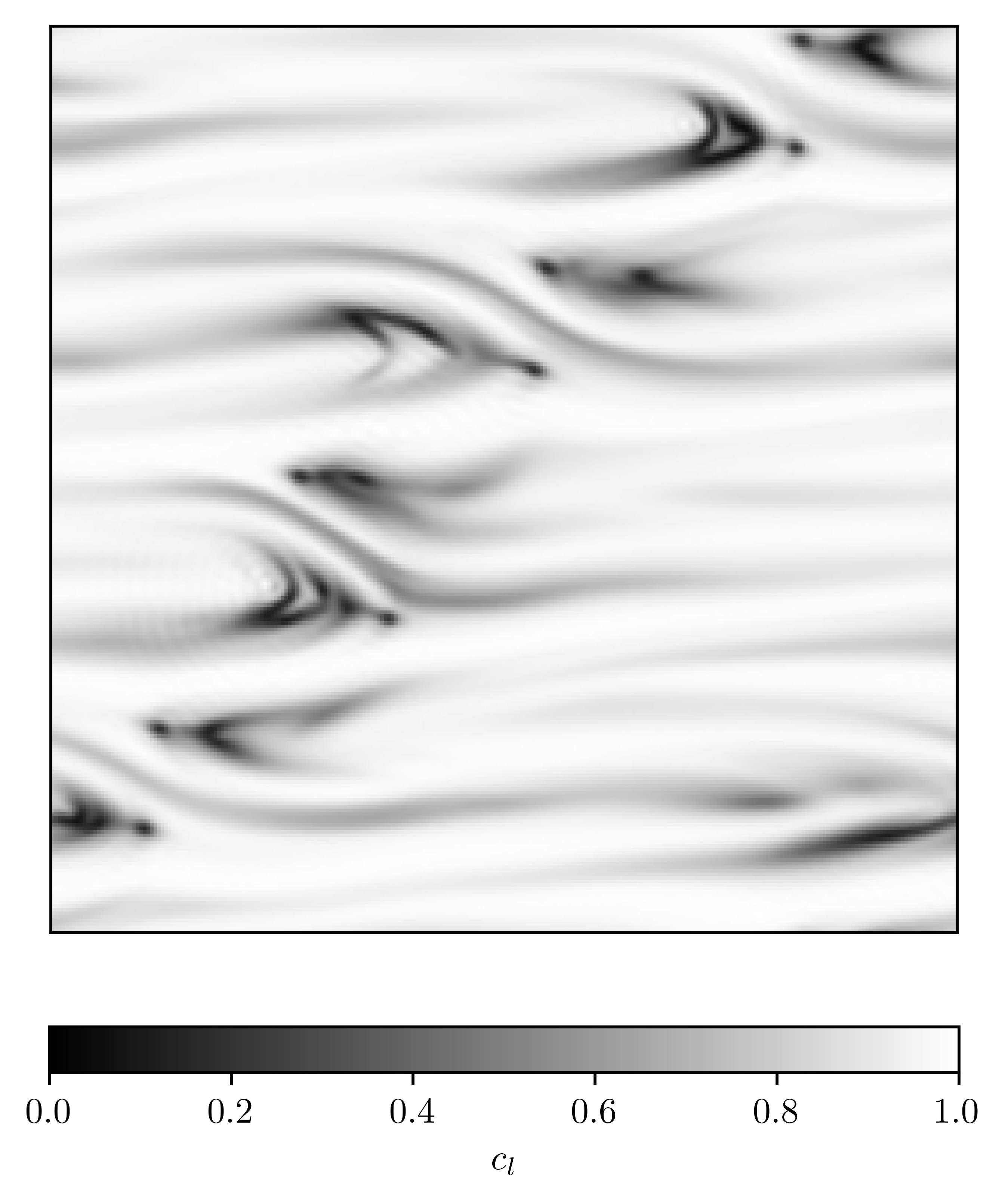}
	\includegraphics[width=0.82\linewidth]{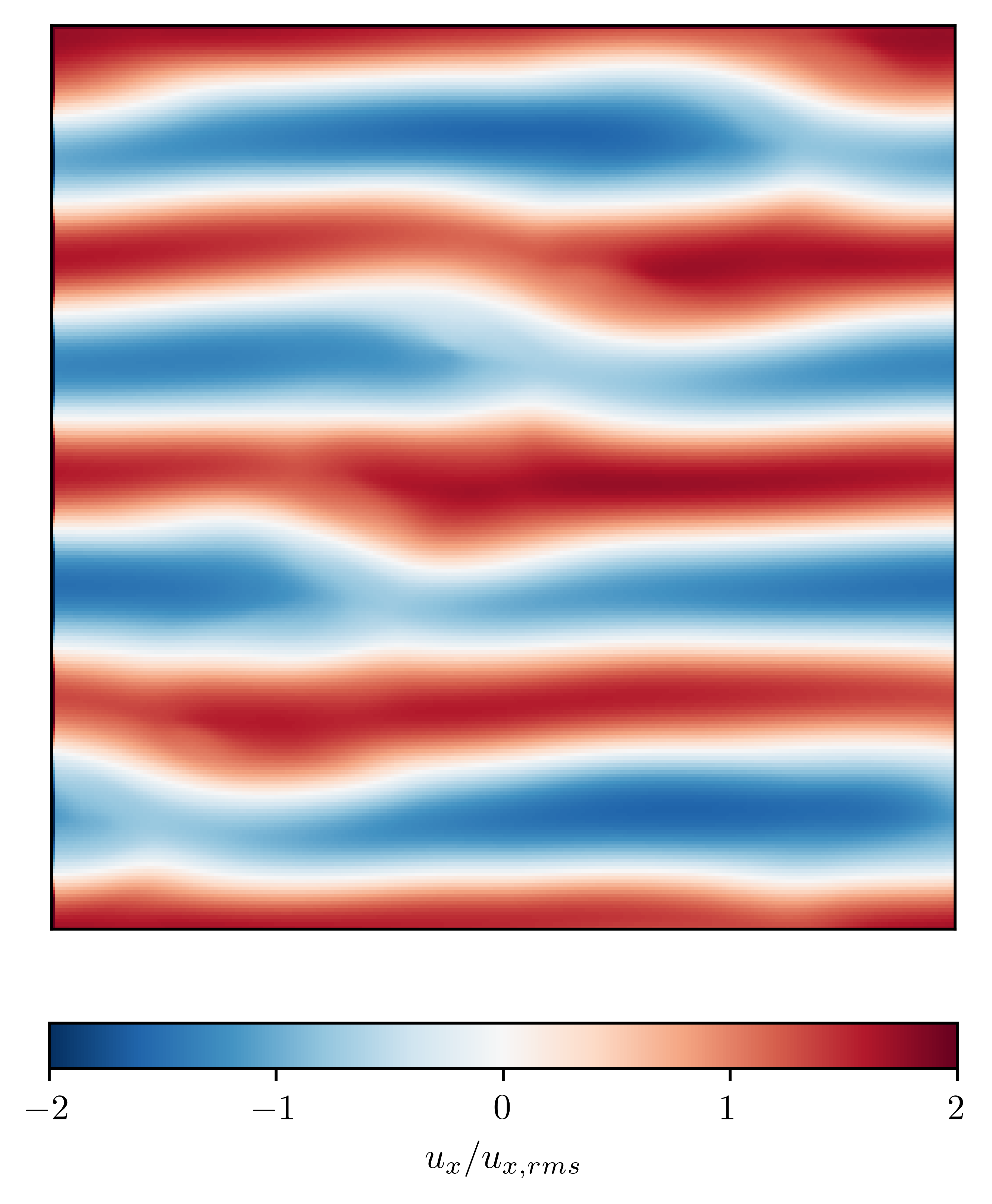}
	\caption{Vertical sections in the $x$-$z$ plane of the uniaxiality parameter $c_l$ (top) and of the streamwise $u_x$ component of the velocity field, normalized with respect to its root-mean-squared value $u_{x,rms}$ (bottom), in the chaotic regime at $\eta_p=7$.
	}
	\label{fig8}
\end{figure}
The pdfs of the isotropy coefficient $c_s$ conditioned to $flow$- and $shear$-dominated regions shows that the points with larger values of $c_s$ are mostly located into the flow-dominated regions: $P^{\left(F\right)}(c_s)$ is almost an order of magnitude larger than $P^{\left(S\right)}(c_s)$ for $c_s \gtrsim 0.4$ (Fig. \ref{fig6}). As expected, isotropic polymer configurations are disfavored in the presence of a mean shear.  

Furthermore, the mean shear influences also the statistics of biaxial regions.
As it is shown in Fig.~\ref{fig7}, the joint pdf of the angles $\{\theta_3, \phi_3\}$,
which describes the orientation of $\bm{l}$
is remarkably different in the flow- and the shear- dominated biaxial regions.
In particular, in the flow-dominated regions $P_B^{(F)}(\theta_3,\phi_3)$ displays a local maximum at $\{ \theta_3 \approx \pi/2, ~\phi_3 \approx 0 \}$, corresponding to biaxial order in the $x$-$z$ plane. This maximum is almost absent in the shear-dominated regions (the value of the pdf is approximately an order of magnitude smaller compared to the flow-dominated ones).
Moreover, in the $shear-$dominated regions the pdf $P_B^{(S)}(\theta_3,\phi_3)$ is more centered around $\{ \theta_3 \approx \pi/2, ~ \phi_3 \approx \pi/2 \}$. 
Therefore, the mean shear suppresses the biaxial configuration,
with the exception of biaxiality in the plane $x$-$y$
which is perpendicular to the direction of the mean shear. 

This phenomenon has a simple explanation. In the Kolmogorov flow, the most intense component of the velocity gradients is $\partial_z u_x$. A biaxial configuration of the polymers in the $x$-$y$ plane (or a generic uniaxial configuration with $\bm{n}$ lying in the $x$-$y$ plane), corresponding to $R_{i3}=0$, minimizes the product $\partial_j u_i R_{ij}$, and therefore the stress exerted by the rods on the fluid.

\subsection{Topological defects and chaotic flow}
The existence of correlations between the chaotic flow
and the topological defects of the configuration tensor
$\bm{R}$ can be qualitatively inferred by the comparison of a section of the streamwise component
of the velocity $u_x$ with the corresponding section of the uniaxiality parameter $c_l$ (Fig. \ref{fig8}).
We clearly observe that the regions with $c_l \ll 1$, \textit{i.e.} topological defects of the director field,
are associated with deformations of the mean flow (see also the Movie 1 in the Supplementary Material).

In order to highlight these correlations, 
it is convenient to decompose the velocity field $\bm{u}\left( \bm{x},t \right)$
as the sum of the mean flow and the fluctuations field:
\begin{equation}
\bm{u}\left(\bm{x}, t \right) = \overline{u_x}\left( z\right)\hat{\bm{e}}_x + \bm{u}'\left(\bm{x},t \right).
\end{equation}
\begin{figure}[h!]
	\centering
	\includegraphics[width=0.81\linewidth]{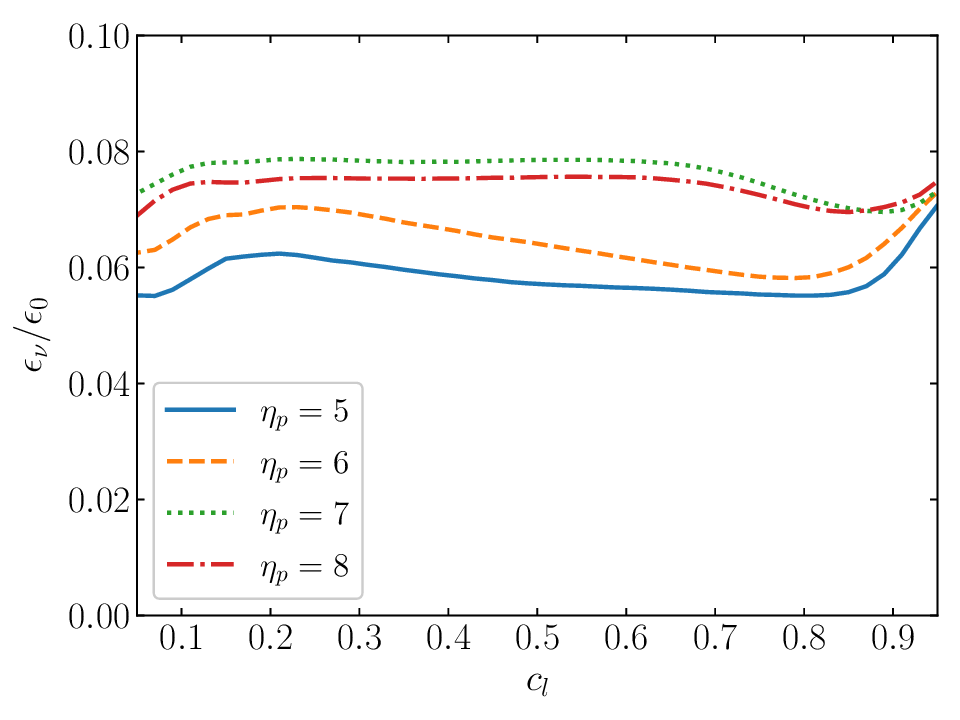}
	\includegraphics[width=0.81\linewidth]{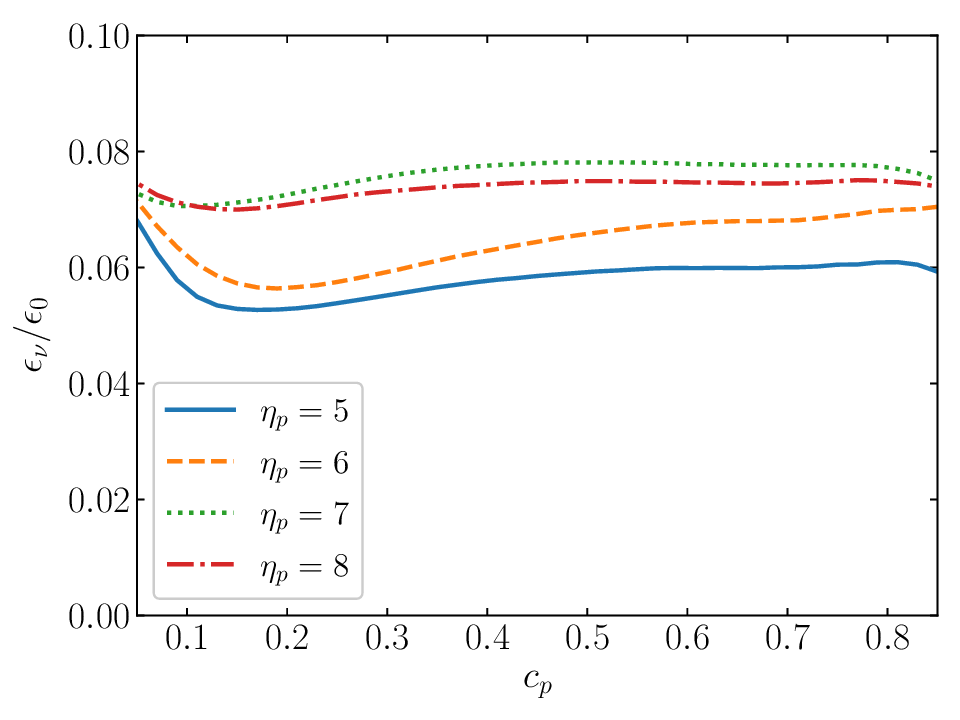}
	\includegraphics[width=0.81\linewidth]{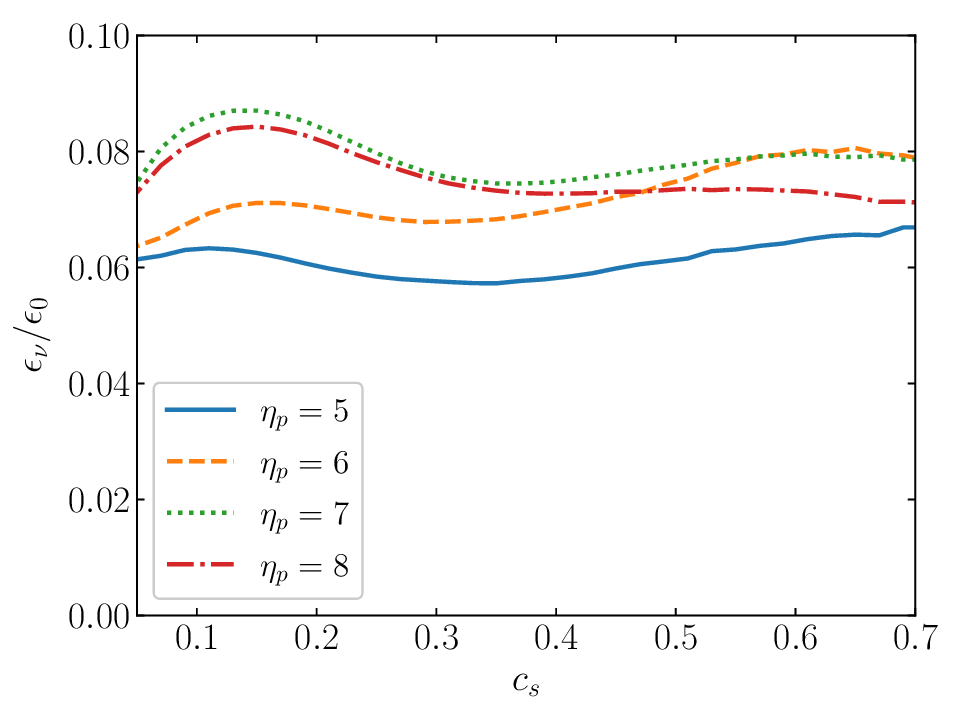}
	\caption{Local dissipation of kinetic energy due to viscosity
          $\epsilon_{\nu} = \nu | \pmb{\nabla} \times\bm{u}' |^2$
          normalized with $\epsilon_0 = FU_0/2$
          as a function of uniaxiality coefficient $c_l$ (top),
          biaxiality coefficient $c_p$ (center) and isotropy coefficient $c_s$ (bottom).
	}
	\label{fig9}
\end{figure}
From the velocity fluctuation field $\bm{u}'$ we 
define the vorticity fluctuations field $\pmb{\omega}' = \pmb{\nabla} \times \bm{u}'$ 
as well as the fluctuating part of the polymer stress tensor
$\sigma_{ij}' = 6 \nu \eta_p R_{ij} \left( \partial_l u'_k \right) R_{kl}$.
The global budget of the kinetic energy associated with the velocity fluctuations
$ 1/2 \langle |\bm{u}'|^2 \rangle$ can be obtained from equation \ref{eq:sys1a}:
\begin{equation}\label{eq:2a}
\frac{1}{2} \frac{d}{dt} \langle |\bm{u}'|^2 \rangle   = -\nu \left\langle |\pmb{\omega}'|^2 \right\rangle - \langle \sigma'_{ij} \partial_j u'_i\rangle,
\end{equation}
where the angular brackets $\langle \cdot \rangle$ denote the average over the whole volume.
\begin{figure}[h!]
	\centering
	\includegraphics[width=0.81\linewidth]{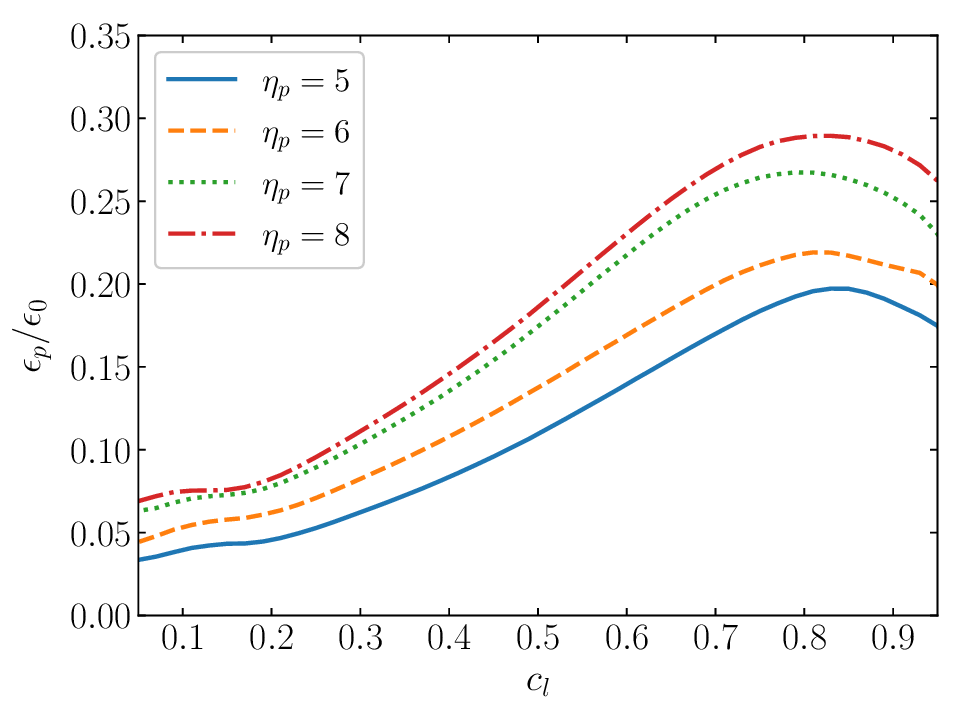}
	\includegraphics[width=0.81\linewidth]{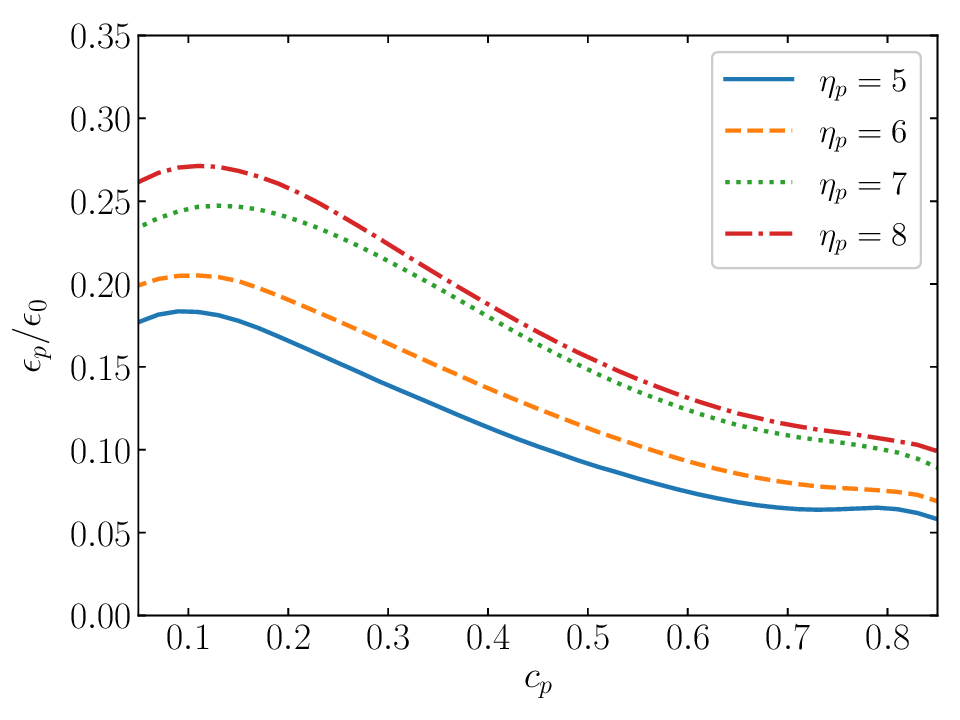}
	\includegraphics[width=0.81\linewidth]{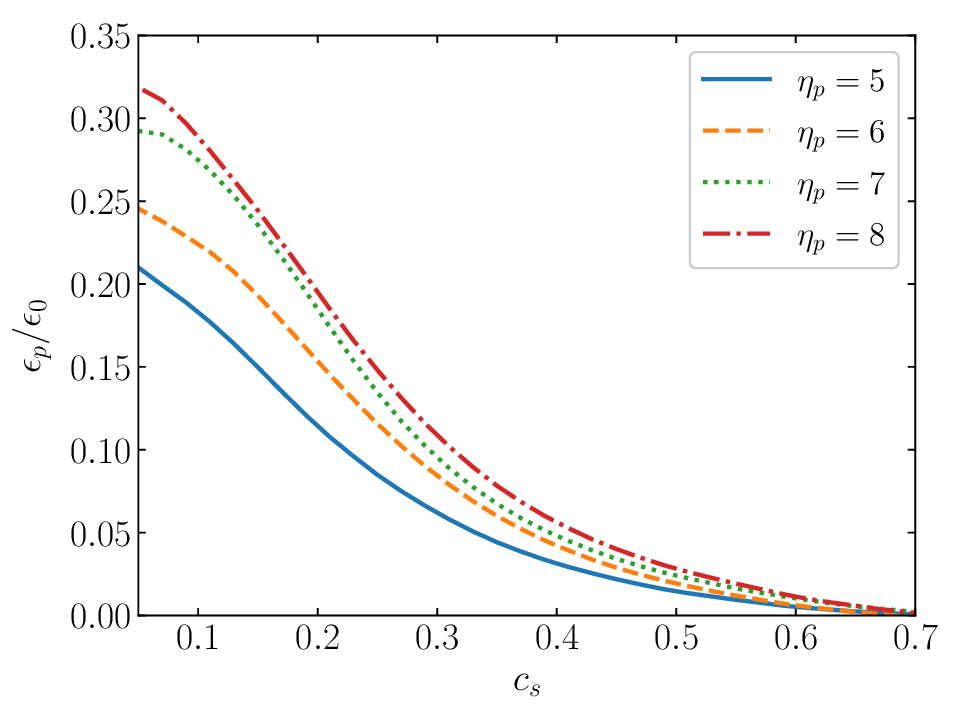}
	\caption{Local dissipation of kinetic energy due to polymer stress $\epsilon_{p} =\sigma'_{ij} \partial_j u'_i$, normalized with $\epsilon_0 = FU_0/2$, as a function of uniaxiality coefficient $c_l$ (top), biaxiality coefficient $c_p$  (center) and isotropy coefficient $c_s$ (bottom).
	}
	\label{fig10}
\end{figure}
Similarly, by taking the curl of Eq. \ref{eq:sys1a} and multiplying it scalarly by $\pmb{\omega}'$, we obtain the global budget of the enstrophy associated with the vorticity fluctuations $1/2 \left\langle |\pmb{\omega}'|^2 \right\rangle $:
\begin{subequations}
\begin{align}\label{eq:2b}
  \frac{1}{2}\frac{d}{dt} \left\langle |\pmb{\omega}'|^2 \right\rangle =
  - \langle \left( \partial_j \omega'_i \right) u'_i \omega'_j \rangle
  - \nu \left\langle  | \pmb{\nabla} \times \pmb{\omega}'|^2 \right\rangle \nonumber \\
  - \langle \left( \partial_i \omega'_j \right) \left( \varepsilon_{jkl} \partial_k \sigma'_{il} \right) \rangle,
\end{align}
\end{subequations}
where $\varepsilon_{ijk}$ is the three-dimensional Levi-Civita symbol. Eqs.~\eqref{eq:2a} and \eqref{eq:2b} have been obtained taking explicitly into account the periodic boundary conditions of the domain.

In order to quantify the relationship of the velocity and vorticity fluctuations
with the rods orientation we compute the averages of the various terms which contribute
to the energy and enstrophy dissipation
$\epsilon_{\nu} = \nu | \pmb{\nabla} \times\bm{u}' |^2$ (viscous energy dissipation) 
$\epsilon_p = \sigma'_{ij} \partial_j u'_i$ (polymer energy dissipation)
$\zeta_{\nu} = \nu |\pmb{\nabla} \times \pmb{\omega}'|^2 $ (viscous enstrophy dissipation)
and
$\zeta_p = \left( \partial_i \omega'_j \right) \left( \varepsilon_{jkl} \partial_k \sigma'_{il} \right)$
(polymer enstrophy dissipation)
conditioned to the local values of $c_l$, $c_p$ and $c_s$.
\begin{figure}[h]
	\centering
	\includegraphics[width=0.81\linewidth]{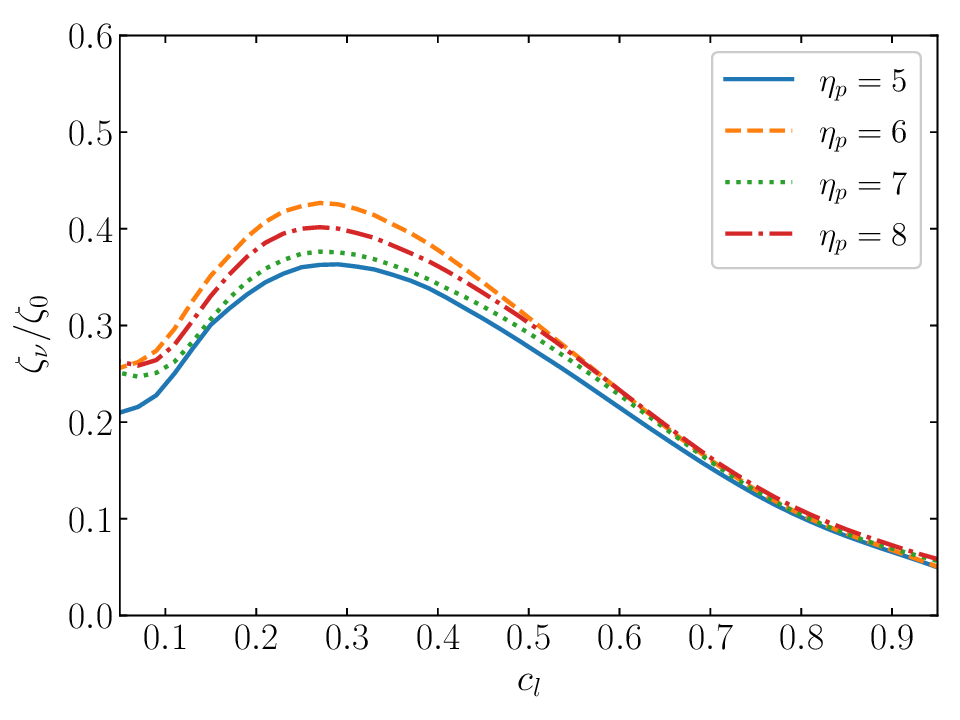}
	\includegraphics[width=0.81\linewidth]{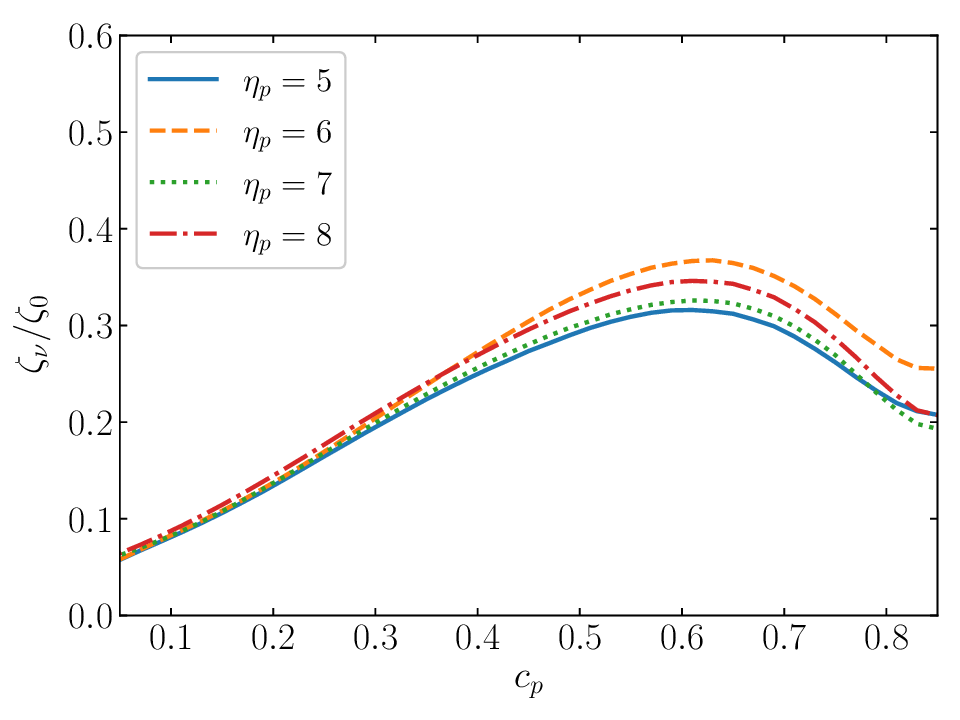}
	\includegraphics[width=0.81\linewidth]{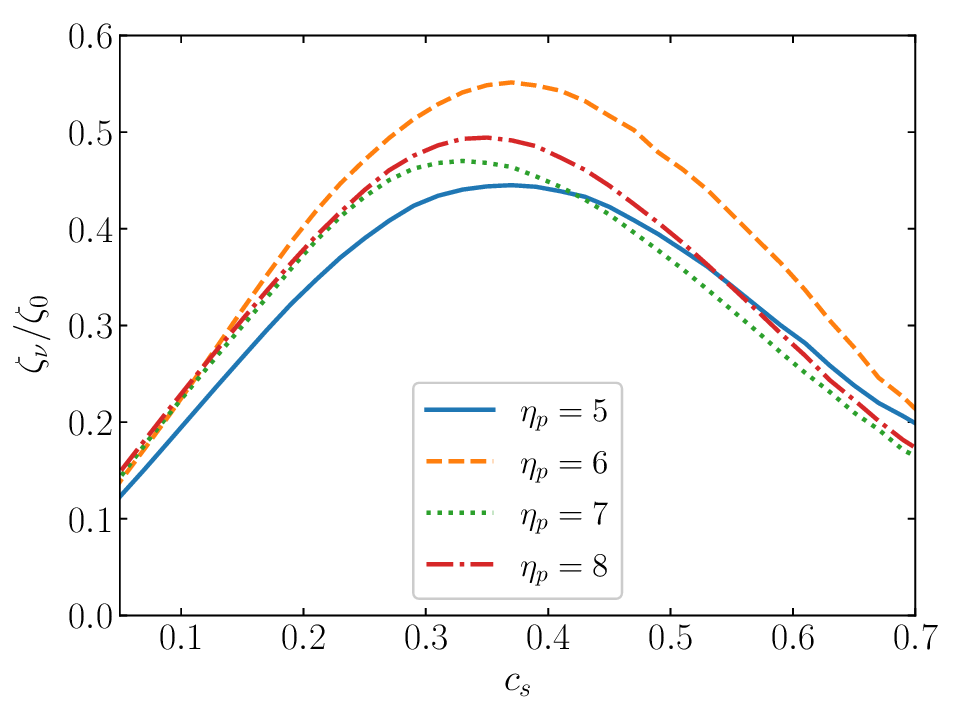}
	\caption{Local dissipation of enstrophy due to viscosity
		$\zeta_{\nu} = \nu |\pmb{\nabla} \times \pmb{\omega}'|^2 $,
		normalized with $\zeta_0 = FU_0K^2/2$, as a function of uniaxiality coefficient $c_l$ (top),
		biaxiality coefficient $c_p$ (center)
		and isotropy coefficient $c_s$ (bottom).
	}
	\label{fig11}
\end{figure}
The \textit{vortex stretching} term
$\langle \left( \partial_j \omega'_i \right) u'_i \omega'_j \rangle$
in Eq.~\eqref{eq:2b} is negligible because
of the low Reynolds number of the chaotic flow\cite{puggioni2022enhancement}. 
\begin{figure}[h]
	\centering
	\includegraphics[width=0.81\linewidth]{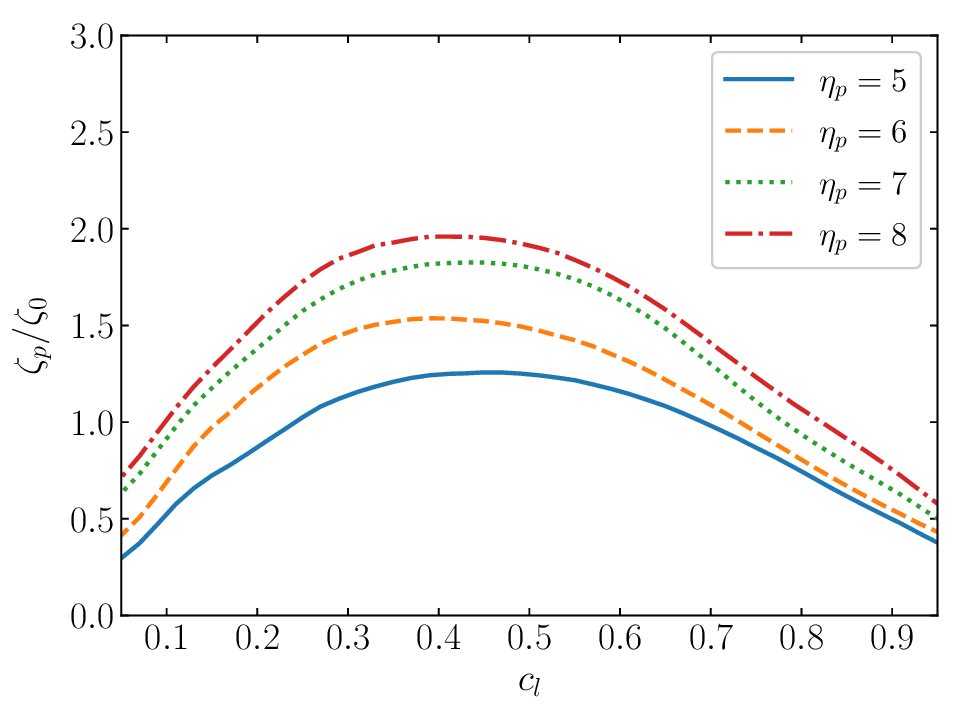}
	\includegraphics[width=0.81\linewidth]{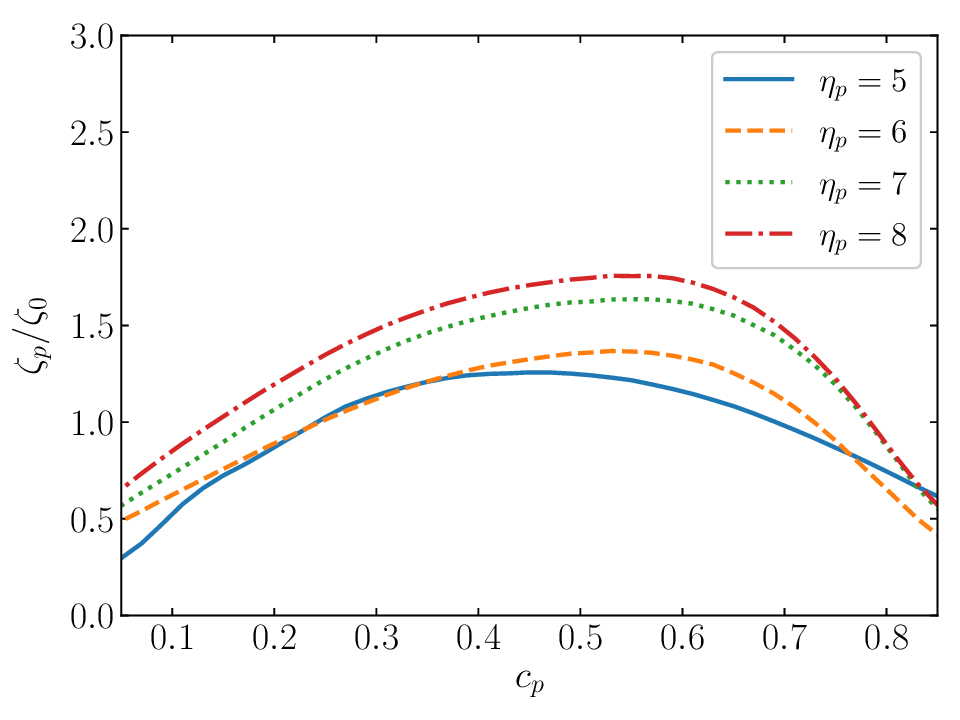}
	\includegraphics[width=0.81\linewidth]{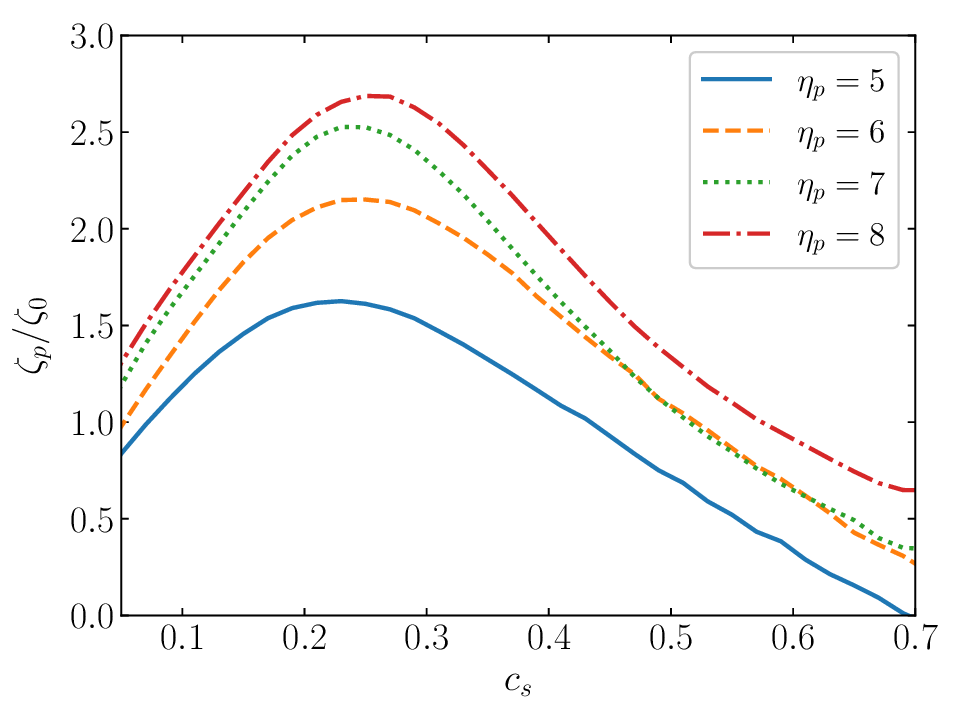}
	\caption{Local dissipation of enstrophy due to polymer stress
          $\zeta_p = \left( \partial_i \omega'_j \right) \left( \varepsilon_{jkl} \partial_k \sigma'_{il} \right)$,
          normalized with $\zeta_0 = FU_0K^2/2$,
          as a function of uniaxiality coefficient $c_l$ (top),
          biaxiality coefficient $c_p$ (center) and isotropy coefficient $c_s$ (bottom).
	}
	\label{fig12}
\end{figure}

Figs. \ref{fig9} and \ref{fig10} show the conditional
average of the contributions $\epsilon_{\nu}$ and $\epsilon_p$ of the energy dissipation,
normalized with $\epsilon_0 = FU_0/2$.
The viscous dissipation $\epsilon_{\nu}$ is almost independent on $c_l, c_p, c_s$ (see Fig.\ref{fig9}),
meaning that the intensity of velocity gradients is not significantly influenced by the local rotational order.
Conversely, the energy dissipation due to the polymers is larger in uniaxial regions,
as signaled by the conditional average of $\epsilon_p$ which attains a maximum
for $c_l \approx 0.8$, $c_p \approx 0.1$ and $c_s \approx 0$ (see Fig.\ref{fig10}).
Moreover, it displays a clear monotonic dependence on the polymer concentration $\eta_p$.  
This phenomenon has a simple explanation. 
Recalling that $\sigma'_{ij} = 6 \nu \eta_p R_{ij} \left( \partial_l u'_k \right) R_{kl}$
and that $\epsilon_p =  \sigma'_{ij} \partial_j u'_i$
we get that $\epsilon_p$ has a quadratic dependence with respect to the velocity gradients
and to the configuration tensor.\cite{doi1988theory}
Given that the intensity of the velocity gradients is not significantly influenced by the orientation of the polymer
(as suggested by the conditional average of $\epsilon_{\nu}$),
the intensity of $\epsilon_p$ can be estimated by assuming a diagonal form of $R_{ij}$,
which gives $\epsilon_{p} \propto \text{Tr} \bm{R}^2$.
The trace $\text{Tr}\bm{R}^2$ is precisely maximum in uniaxial regions.
The physical meaning of this result is that the energy dissipation due to the polymer stress is maximum
in the regions in which the polymer orientation displays uniaxial order. 

The conditional average of the dissipative terms of the enstrophy balance is shown in Figs. \ref{fig11} and \ref{fig12}.
The enstrophy dissipation due to viscosity $\zeta_{\nu}$ is maximum for
$c_l\approx 0.3$, $c_p \approx 0.7$ and $c_s \approx 0.4$, and it is much larger that the value for $c_l \approx 0.95$
(see Fig.~\ref{fig11}). 
This results shows that strong vorticity gradients are associated with the topological defects
(i.e., regions in which $c_l \ll 1$).
In particular, we note that the maximum value of the conditional average of $\zeta_{\nu}$
occurs for values of $c_l$ and $c_p$ which are close, but not exactly equal to a perfect biaxial configuration
$c_l \approx 0$ or $c_p \approx 1$. We can therefore infer that the regions in which the vorticity gradients
are more intense correspond to the boundaries of the biaxial regions.
The conditional average of $\zeta_p$ displays a qualitatively similar behavior (see Fig.~\ref{fig12}) 
but with a smaller ratios between the maximum and the minimum values.
This is the results of a competition between the dependence of $\zeta_p$ on the intensity of the vorticity gradients,
which is maxima close to the regions of prevalent biaxial order,
and its dependence on $\text{Tr} \bm{R}^2$ which is maximum in the uniaxial regions. 
It is important to remark that the term
$\zeta_p = \langle \left( \partial_i \omega_j \right) \left( \varepsilon_{jkl} \partial_k \sigma'_{il} \right) \rangle$
is not a priori necessarily dissipative.

\subsection{The two-dimensional model}

Previous studies showed that the chaotic regime with rodlike polymers appears to be qualitatively
very similar between 2D and 3D simulations,\cite{puggioni2022enhancement}.
Therefore it is natural to investigate the interplay between the solvent velocity field and the
rotational order of the rods also in 2D flow.
In this simplified geometry, the uniaxiality coefficient $c_l=\lambda_1-\lambda_2 = S$
is sufficient to characterize the local degree of order,
while a single angle $\theta = \arccos |n_x|$ describes the orientation of the director
$\bm{n}$ with respect to the direction of the mean flow.

\begin{figure}[h]
	\centering
	\includegraphics[width=0.83\linewidth]{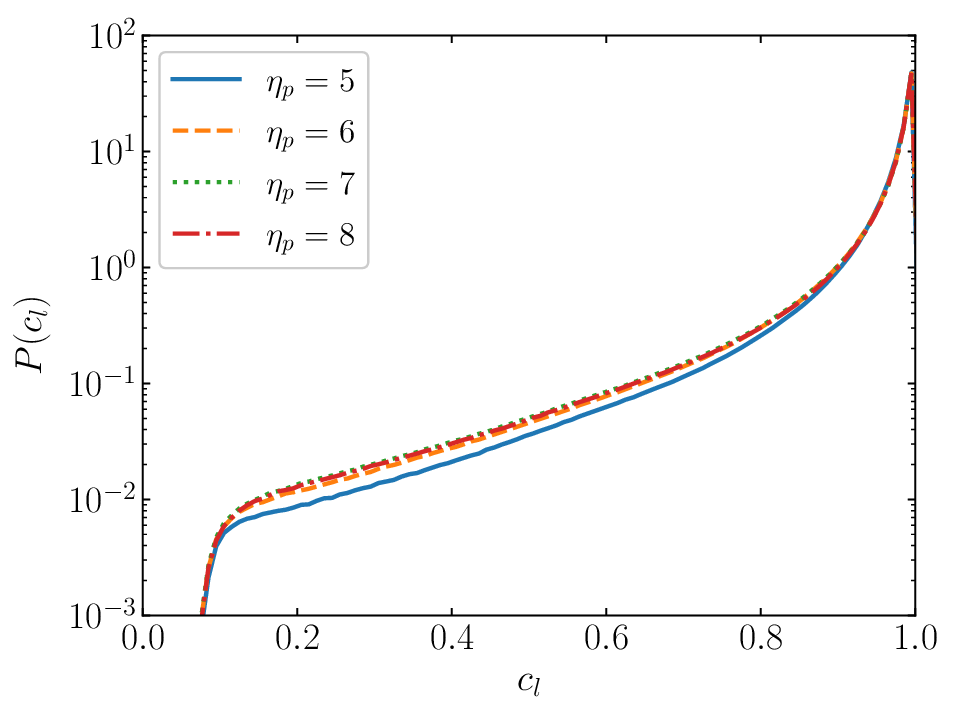}
	\quad
	\includegraphics[width=0.83\linewidth]{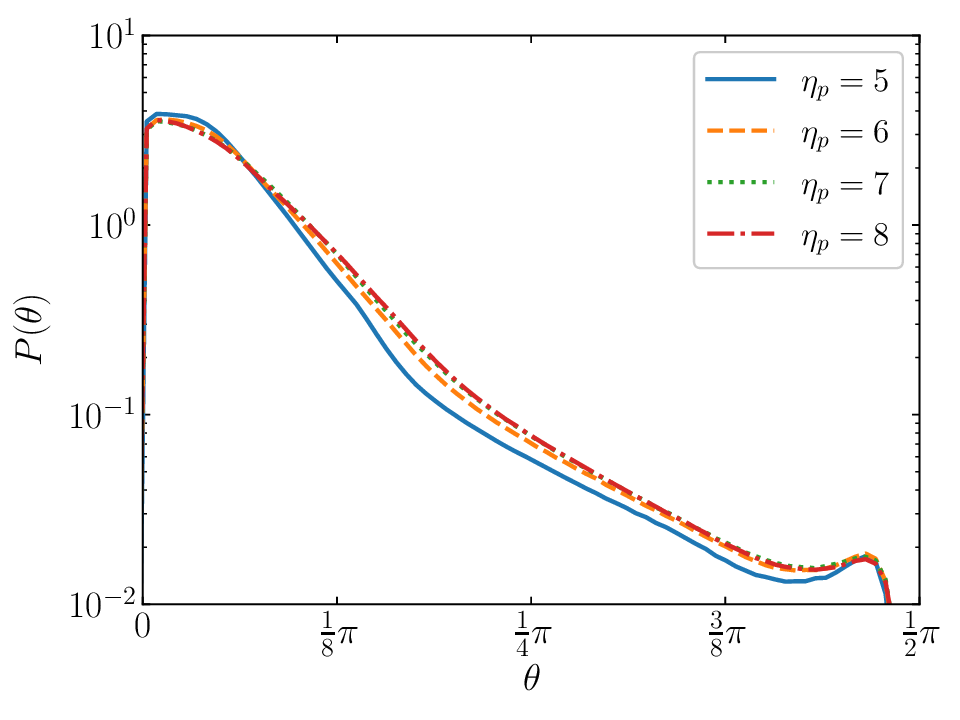}
	\caption{ Top: pdfs of the uniaxiality coefficient $c_l$ in 2D simulations, for different values of concentration coefficient $\eta_p$. Bottom: pdfs of the angle $\theta$ between the director $\bm{n}$ and the $x$-axis, for different values of concentration coefficient $\eta_p$.
	}
	\label{fig13}
\end{figure}

The statistics shown in Fig. \ref{fig13} is similar to the results of the 3D simulations:
in the vast majority of spatial locations the polymers display uniaxial order ($c_l \approx 1$),
and they are mostly aligned in the direction of the mean flow ($\theta \approx 0$).
The absence of a spanwise direction, and therefore of the impossibility to have biaxial order in a plane perpendicular
to the mean shear, causes the points deviating from the uniaxial order to be significantly reduced compared to the 3D system.
This is due to the fact that, in the presence of a strong transverse shear in 2D flow, 
the only configuration that minimizes the product $\partial_j u_i R_{ij}$ (and hence the polymer stress) 
is the uniaxial one with $\bm{n}$ parallel to the $x$-axis ($c_l=1$, $\theta = 0$). 

\begin{figure}[h]
	\centering
	\includegraphics[width=0.83\linewidth]{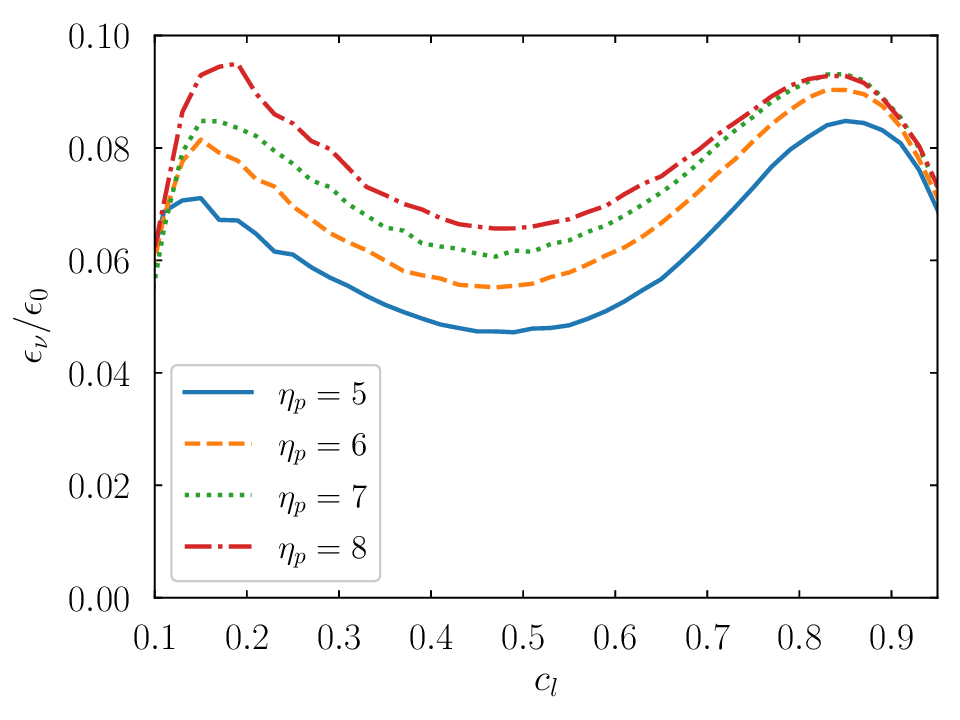}
	\quad
	\includegraphics[width=0.83\linewidth]{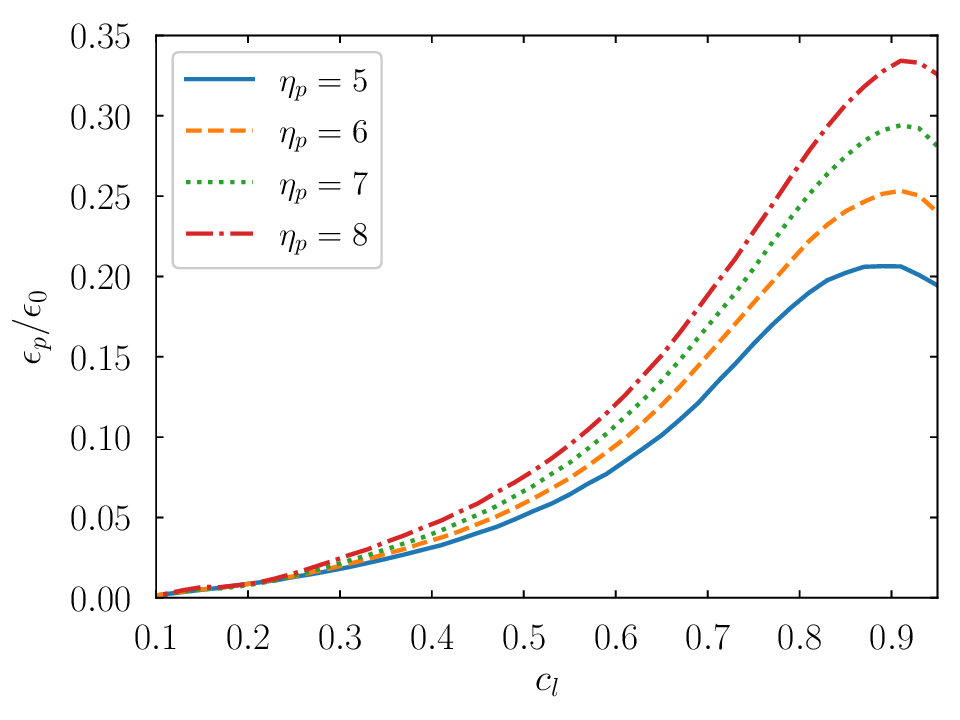}
	\caption{ Dissipation of kinetic energy due to viscosity
          $\epsilon_{\nu} =  \nu | \pmb{\nabla} \times\bm{u}' |^2$
          (top) and due to polymer stress $\epsilon_{p}= \sigma'_{ij} \partial_j u'_i$ (bottom),
          normalized with $\epsilon_0 = FU_0/2$,
          as a function of the uniaxiality coefficient $c_l$,
          for different values of concentration coefficient $\eta_p$,
          in 2D simulations.
	}
	\label{fig14}
\end{figure}
\begin{figure}[h]
	\centering
	\includegraphics[width=0.83\linewidth]{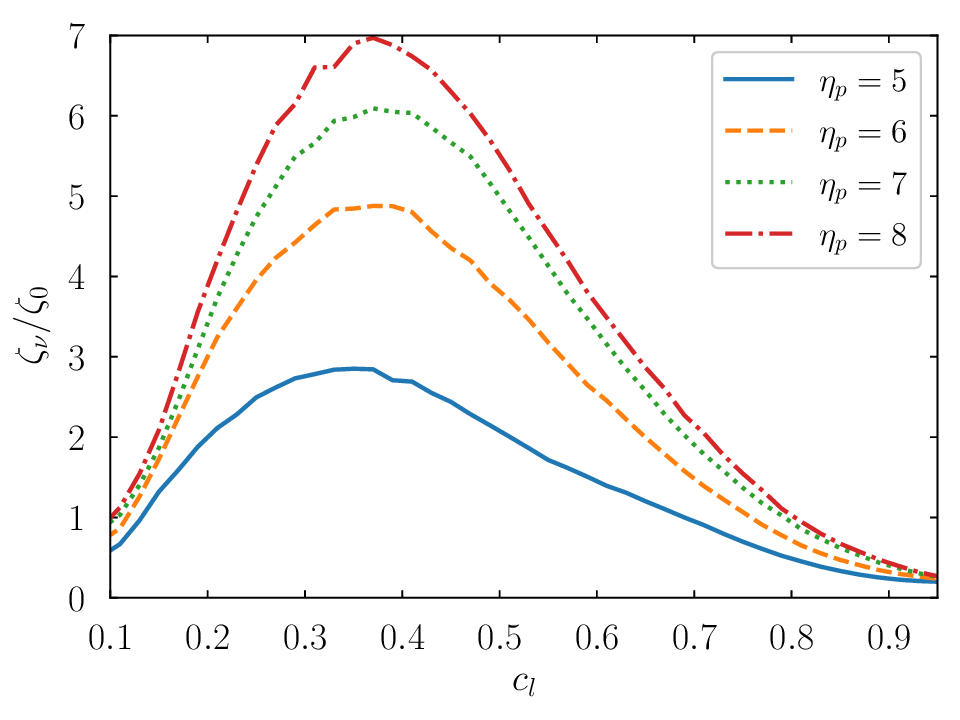}
	\quad
	\includegraphics[width=0.83\linewidth]{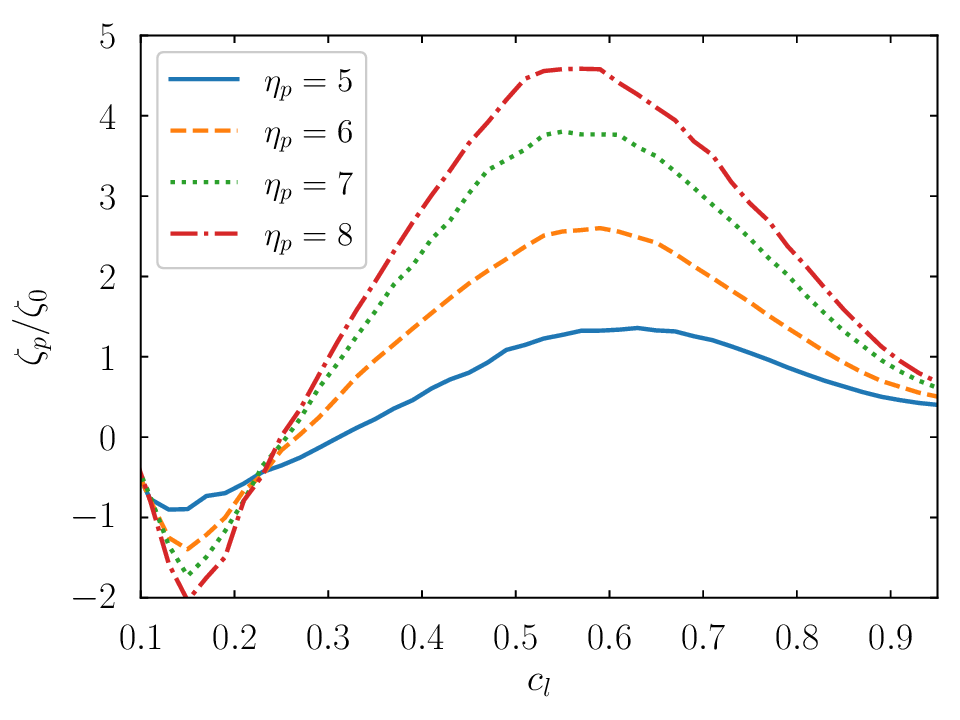}
	\caption{ Dissipation of enstrophy due to viscosity
          $\zeta_{\nu} = \nu |\pmb{\nabla} \times \pmb{\omega}'|^2 $ (top)
          and due to polymer stress
          $\zeta_{p}=  \left( \partial_i \omega'_j \right) \left( \varepsilon_{jkl} \partial_k \sigma'_{il} \right)$ (bottom),
          normalized with $\zeta_0 = FU_0K^2/2$,
          as a function of the uniaxiality coefficient $c_l$,
          for different values of concentration coefficient $\eta_p$,
          in 2D simulations.
	}
	\label{fig15}
\end{figure}
As for the 3D simulations, we investigated the correlations between the velocity and enstrophy fluctuations
and the polymer orientational order by computing the conditional average
of the local terms $\epsilon_{\nu}$, $\epsilon_p$, $\zeta_{\nu}$ and $\zeta_p$ (defined as in the 3D case)
which contribute to the energy and enstrophy balance. 
The behavior of the conditional average of $\epsilon_{\nu}$, $\epsilon_p$ observed in 2D simulations
is qualitatively similar to the 2D system (see Fig. \ref{fig14}). 
The viscous dissipation $\epsilon_{\nu}$ is weakly affected by the degree of order of the polymers, 
although the dependence on concentration $\eta_p$ is clearly observable.
We remind that in the 2D simulations the concentration parameter $\eta_p$ has been rescaled
according to the dimensional relation $\eta_p^{2D} = (2/3) \eta_p^{3D}$\cite{puggioni2022enhancement}.
To facilitate the comparison, we report the 3D-equivalent value in the figures.

The dissipation due to the polymers $\epsilon_p$ is more intense in uniaxial regions,
as shown by the maximum of $\epsilon_p$ close to $c_l \approx 0.9$.  
For the terms related to the enstrophy balance
we find that $\zeta_{\nu}$ is maximum in regions with $c_l \approx 0.4$,
which do not displays neither an uniaxial order ($c_l =1$)
nor a perfect 2D isotropic order ($c_l = 0$) (see Fig. \ref{fig15}).
This result shows that the regions surrounding the topological defects
are characterized by strong vorticity gradients.

Comparing the 2D and 3D cases, we observe that in 2D flow
the dependence of $\zeta_{\nu}$ on $c_l$ is more pronounced,
as well as the dependence on the concentration $\eta_p$.
The increased intensity of vorticity gradients around the topological defects in 2D with respect to the 3D system
is easily explained in terms of the different stress associated to the topological defects. 
In 3D the majority of the defects has a biaxial order in the $x$-$y$ plane, perpendicular to the mean shear.
This configuration which minimizes the polymer stress is not allowed in 2D.
The topological defects in 2D display biaxial order in the $x$-$z$ plane, causing a strong local stress.  
We also notice that the conditional average of the polymer contribution to the enstrophy balance $\zeta_p$
becomes negative in regions of strong rotational disorder $c_l \lesssim 0.2$.
In these regions, the rotational dynamics of polymers acts on average as a source of enstrophy in 2D, 
a phenomenon which is not observed in the 3D simulations. 

\section{Conclusions}\label{sec:con}

In this work, we presented the
results of direct numerical simulations of the Doi-Edwards model
for a dilute solution of inertialess rodlike polymers,
aimed to study the interplay between the microscopic rotational order of the polymer phase
and the chaotic low-Reynolds flow generated by the polymer stress.
The simulations have been carried out both in 3D and 2D flows,
allowing to investigate the role of the dimensionality of the system.  

Our findings highlight the role of the topological defects in this "turbulent-like" regime at low $Re$. 
In the 3D simulations, we showed that the rods are preferentially aligned with the mean flow,
while the topological defects of the director field take the form
of local regions with biaxial order in the plane perpendicular to the mean shear.
This configuration minimizes the friction between the rods and the fluid and the polymer stress. 

We also investigated the correlation between the rotational order of the polymer
and the velocity and vorticity fluctuations of the chaotic flow,
showing that that maxima of the energy dissipation due to the polymer stress   
occur in regions of the flow characterized by uniaxial order,  
while intense vorticity gradients are observed in the regions surrounding the topological defects.

The comparison between 3D and 2D simulations reveals a qualitatively similar scenario, with some differences.
In the 2D systems, the impossibility to have biaxial order in the plane orthogonal to the shear, 
minimizing the polymer stress, leads to a reduction of the area covered by the defects
and to a strong increase of the vorticity gradients around them, compared to the 3D system.

In conclusion, the picture which emerges from our results is that
while the topological defects are advected by the mean flow, they in turn generate disturbance flows around them.
These disturbances cause hydrodynamical interaction within the defects, which ultimately results
in the emergence of the chaotic behavior of the fluid (see also the Movie 1 in the Supplementary Material). 

These results, and in particular the importance of the topological defects in the average polymer orientation, are reminiscent of active nematic turbulence,\cite{alert2022active} which is explained in terms of the interplay between the topological defects of the  director field and the geometry of the flow, both in two\cite{giomi2015geometry,doostmohammadi2018active,brezin2022spontaneous} and in three\cite{vcopar2019topology,duclos2020topological,krajnik2020spectral,kralj2023defect} spatial dimensions. For these reasons, further investigations on the dynamics of the chaotic regime generated by rodlike polymers could be carried out adopting similar techniques to those for active nematics, such as the analytical computation of the backflow induced by the defects.\cite{giomi2014defect,thampi2014instabilities,brezin2022spontaneous}


\section*{Acknowledgments}
We acknowledge HPC CINECA for computing resources (INFN-CINECA grant no. INFN23-FieldTurb). L.P. acknowledges support from research project TurboPuff, funded by Fondazione CRT.


\bibliography{biblio}

\end{document}